\documentstyle[epsf,graphicx]{elsart}
\begin{document}
\begin{frontmatter}
\title{ Approximate  particle number projection with density
dependent forces: Superdeformed bands in the A=150 and A=190 regions}
\author{ A. Valor\thanksref{kk}}, 
\author{ J.L. Egido and L.M. Robledo }
\address{Departamento de F\'{\i}sica Te\'orica C-XI\\
Universidad Aut\'onoma de Madrid, E--28049 Madrid, Spain}
\thanks[kk]{Present address~: Service de Physique Nucl\'eaire Th\'eorique,
       U.L.B.-C.P.229, B-1050 Brussels, Belgium}
\date{\today}
\maketitle

\begin{abstract}
We derive the equations for approximate particle number projection based
on  mean
field wave functions with finite range density dependent forces.
As an application  ground  bands of even-A superdeformed 
nuclei in the $A=150$ and $A=190$ regions are calculated with the Gogny force.
 We discuss nuclear properties such as quadrupole moments, moments of inertia 
and quasiparticle spectra,  among others, as a function of the angular momentum.
We obtain a good overall description.
\end{abstract}
\end{frontmatter}

\section{Introduction}

 Understanding pairing correlations has  been and still remains a challenge for
  nuclear physicists.
Several decades ago it was thought that pairing correlations will only affect  a
few nuclei and the properties of a few observables at low energy and small
angular momentum; mainly the level density of odd nuclei as compared with the
even ones, the energy of the two quasiparticle states and the moments of inertia
of rotational nuclei.
Such simplistic view has changed over the years in such a way that nowadays
we expect pairing correlations playing a major role in many nuclear phenomena. 
To mention a few recent examples  let us remember the important role of pairing
correlations in halo nuclei, in $N=Z$ systems and in the understanding of
the second moment of inertia in superdeformed nuclei.

In spite of all that the theoretical understanding   of the pairing force
is rather limited.
The recognition of this fact has motivated in the last years a proliferation
of approximate methods (Lipkin-Nogami, Kamlah at second order and  exact
particle number projection) and new pairing forces (density dependent, 
surface and volume pairing).

 Within the mean field approaches  the most successful calculations, aiming
 at describing low energy nuclear structure phenomena have been
performed with density dependent forces. The most popular, because
of its simplicity, is the  Skyrme force. Most of the calculations have been
 performed with the SkM$^*$ parameterization \cite{BONCHE,FLOCARD}.
  The main disadvantage of the Skyrme interaction
is  that realistic pairing matrix elements cannot be extracted from the force,
and therefore  the pairing
interaction (normally a monopole pairing term) has to be added by hand.
Recently, other alternative forms of the pairing interaction, like density dependent
pairing have been explored in connection with the Skyrme interaction 
\cite{THB.95,HDN.98,RBF.99}.
Theories beyond mean field with simple pairing forces have been also 
developed \cite{CHA.93}.
Pairing properties, on the other hand, are very important in the description
of high spin properties. Since the pioneer work of Mottelson and Valatin
\cite{MOVA} many papers and experiments have been devoted to this topic.
This is why the investigation of high spin states with a
realistic force  that is able to treat the short range correlations properly is most
desirable. 
The Gogny force \cite{GOG.75} is well known for providing, with the same set
of parameters, a good description of many nuclear properties at low spins
over the nuclide chart as, for example, ground state energies, odd-even
mass differences, electron scattering data \cite{DEG.80}, fission barriers
\cite{BGG.84}, etc. The main ingredients of the force are the
phenomenological density dependent term which was introduced to simulate the
effect of a G-matrix interaction and the finite range of the force which
provides the pairing matrix elements from the force itself.
In recent calculations the Gogny force has been applied to high spin problems
at normal deformations \cite{ER.93} and to superdeformed shapes \cite{ERC.94,GDB.94}.

In these calculations the cranked HFB approximation (CHFB) was used. The angular momentum and
number of particles of the system were determined by the corresponding constraints
on the wave function.
  It turned out \cite{ER.93}, however, that in spite of the finite range of the force,
the condition of holding the particle number to be right on the average 
was not enough and  a superfluid normalfluid sharp phase transition was found,
 contrary to what one would expect for a finite system.  To remedy this problem, 
 in  \cite{VER.96}
 we have formulated the Lipkin-Nogami method  for density dependent forces,
the so-called CHFB Lipkin-Nogami approximation(CHFBLN).
When this formalism was applied the sharp phase transition  was washed out, the results
of the calculations being in much better agreement with the experiment. 

The purpose of this paper is to report on calculations in the $A=190$ and $A=150$
regions within the CHFB and CHFBLN approximations and the Gogny force.
In section 2.1 we first derive the Lipkin-Nogami approximation from the Kamlah expansion
to second order. In sect.~2.2 we extend this formalism to density dependent forces
 obtaining an extra contribution due to the rearrangement term. In sect.~3.1 we
discuss results for the $A=190$ region, namely for the $^{190-192-194}$Hg and
$^{194}$Pb, and in sect.~3.2 the $A=150$ region, taking as examples the $^{152}$Dy and
$^{150}$Gd nuclei. We conclude the paper with conclusions and some appendixes.

\section{Projected Mean Field Theories}

 Mean field theories (MFT) are based on the assumption of a product wave function;
they constitute the basic (and in most  cases the only one feasible) approximation
to a many body system. The success of MFT is based on the concept of spontaneous
symmetry breaking. By this mechanism the Hilbert space is  considerably enlarged
and many correlations can be taken into account. Obviously, the corresponding
wave functions have to be considered as ``intrinsic'' wave functions.
 The most sophisticated mean field theory is the HFB approximation, 
in which the intrinsic wave function $|\Phi  \rangle $ is
defined as the  non-trivial vacuum to the quasi-particle annihilation operators
\begin{equation}  \label{bogtrans}
\alpha _\mu =\sum_iU_{i\mu }^{*}c_i+V_{i\mu }^{*}c_i^{+},
\end{equation}
where $\left\{ c_i,c_i^{+}\right\} $ is the set of single particle creation
and annihilation operators in terms of which the quasi-particle operators
are expanded, $U$ and $V$ are the so-called Bogoliubov wave functions.
 Product wave functions, $|\Phi  \rangle $, of quasiparticles $\alpha_\mu$,  
in general, are neither eigenfunction
of the angular momentum nor of the particle number operator. 
For heavy, well deformed nuclei, the usual approach of demanding that, at least, 
the expectation values of the angular momentum 
and the particle number operators take the right value, is a good approximation for the angular 
momentum but not for the particle number. This has to do with the number of nucleons
participating in the breaking of the associated symmetry \cite{RS.80}.

The ideal treatment of pairing correlations in nuclei is particle
number projection (PNP) before the
variation  \cite{RS.80}. At high spin this theory
is rather complicated to implement and
up to now  it  has only  been applied to  separable forces \cite{EGPNP}.
On the other hand, the semi-classical recipe of solving the mean
field equations with a constraint on the
corresponding symmetry operator can be derived as the first order
approach to a full quantum-mechanical
expansion (the Kamlah expansion \cite{KAM.68}) of the projected
quantities.
The second order in this
expansion takes into account  the particle number fluctuations and might
cure some of the deficiencies of the first order approximation. However,
full selfconsistent calculations up to second order are rather cumbersome and just 
simple model calculations have been carried out up to now \cite{FLO.94}.
Most second order calculations have been done using the Lipkin-Nogami (LN)
recipe, which was originally proposed in Refs.~\cite{LIP.60,NOG.64,GN.66}.
Recently,  the LN method has been applied to study
superdeformed nuclei at high spins
adding a monopole pairing interaction to the Woods-Saxon
plus Strutinsky approximation \cite{MCD.93}, and to the Skyrme force \cite{GBD.94}.

The usual derivation of the LN method is not valid for density dependent forces
because the dependence in density is not taken explicitly into account. 
In refs.\cite{VER.96,VER.97} the LN approach was generalized  to  properly deal
with this situation. The
derivation is based in the Kamlah expansion for the particle number projected energy.
 In the new formulation
additional terms arise in the equations determining the
parameters of the theory.  To introduce the approximation and the corresponding notation
we make in the following a short derivation of it. 

\subsection{ First and second order approximation to exact particle number
projection~: Derivation for non-density dependent forces}
 
 The LN method can be derived using different arguments. As the usual ways of 
deriving the LN method are not well suited for a generalization to the case of density
dependent forces, in this section we shall derive the equations for non-density dependent
forces using the Kamlah expansion. This derivation can be easily extended to the density
dependent case, see sec. 2.2.
 
   Let $|\Phi \rangle$ be a product wave function of the Hartree-Fock-Bogoliubov
type, i.e. a particle number symmetry violating wave function. We can generate
an eigenstate of the particle number, $|\Psi_{N}\rangle$, by the projection 
technique \cite{RS.80} 
\begin{equation}
   |\Psi_{N}\rangle = \hat{P}^{N}|\Phi \rangle= \frac{1}{2\pi} \int_{0}^{2\pi}d\phi 
e^{i(\hat{N}- N )\phi}  |\Phi \rangle.
\label{projector}
\end{equation}
   The particle number projected energy is given by
\begin{eqnarray}
  E^{N}_{proj}  =  \frac{\langle\Psi_{N}|\hat{H}|\Psi_{N} \rangle}{\langle\Psi_{N}|\Psi_{N} \rangle}=
    \frac{
   \int_{0}^{2\pi}d\phi e^{-i\phi N}\langle\Phi|\hat{H}e^{i\phi\hat{N}}
   |\Phi\rangle}
   {\int_{0}^{2\pi}d\phi e^{-i\phi N}\langle\Phi|e^{i\phi\hat{N}} |\Phi\rangle} =
    \frac{ \int_{0}^{2\pi}d\phi e^{-i\phi N}h(\phi)}
   {\int_{0}^{2\pi}d\phi e^{-i\phi N}n(\phi)},
 \label{bigeq}
\end{eqnarray}
where we have introduced the Hamiltonian and norm overlaps
\begin{equation}
  h(\phi)=\langle\Phi|\hat{H}e^{i\phi\hat{N}}|\Phi\rangle, \;\;\;\;\;\;\;\;
 n(\phi)=\langle\Phi|e^{i\phi\hat{N}}|\Phi\rangle.
  \label{n(phi)}
\end{equation}
  In the case of large particle numbers and strong deformations in the gauge space associated
to $\hat{N}$, one expects the  $ h(\phi)$ and $n(\phi)$  overlaps to be peaked at $\phi=0$ and to be 
very small elsewhere in such a way that the quotient $ h(\phi)/n(\phi)$ behaves smoothly. 
 In this case one can make an expansion of  $ h(\phi)$ in terms of $n(\phi)$ in the following way \cite{KAM.68}
\begin{equation}
h(\phi)=\sum_{m=0}^{M}h_{m}{\hat{\cal{N}}}^{m}n(\phi),
\label{hphi}
\end{equation}
where we have introduced the Kamlah operator
\begin{equation}
\hat{\cal{N}}= \frac{1}{i}\frac{\partial}{\partial{\phi}}-
   \langle\Phi|\hat{N}|\Phi\rangle, 
\end{equation}   
which is a representation of the particle number operator in the space of the
parameter $\phi$. $M$ represents the order of the expansion, for $M \rightarrow
\infty$ eq.~(\ref{hphi}) is exact. For large symmetry breaking systems and
large particle number one expects convergence with a few terms.
The expansion coefficients $h_m$ are determined  by  applying the operators 
$1, \hat{\cal{N}},..., \hat{\cal{N}}^M$ on eq.~(\ref{hphi}) and taking the 
limit $\phi \rightarrow 0$, one obtains
\begin{equation}
\langle\Phi|\hat{H}(\Delta{\hat{N}})^{n}|\Phi\rangle=
\sum_{m=0}^{M}h_{m}\langle\Phi|(\Delta{\hat{N}})^{m+n}|\Phi\rangle
\label{coefeq}
\end{equation}
with $\Delta\hat{N} = \hat{N} - \langle \Phi | \hat{N}|\Phi \rangle$ and 
$n=0,\dots,M$. The previous equality provides an $M+1$ equations system, 
$h_0,h_1,\dots, h_M$ being the unknowns.
From now on we will use
the shorthand notation $\langle\hat{A}\rangle\equiv\langle\Phi | \hat{A}|\Phi\rangle$.
Substitution of  eq.~(\ref{hphi}) in eq.~(\ref{bigeq}) provides
 the projected energy to order $M$
\begin{equation}
E^{M}_{proj}=\sum_{m=0}^{M}h_{m}(N-\langle\hat{N}\rangle)^{m}
\label{energyM}
\end{equation}
This expression tell us that to determine the wave function $|\Phi\rangle$, in
an approximate (to order $M$) projection before the variation, we have to minimize
eq.~(\ref{energyM}).
To first order, i.e. $M=1$,
one obtains
\begin{equation}
E^{(1)}_{proj}\;=\;h_0\;\;+\;\; h_{1}\;(N - \langle\hat{N}\rangle), 
\label{energy1}
\end{equation}
with
\begin{equation}
 h_0 = \langle\hat{H}\rangle,\;\;\;\;\;\;\;\; h_1= \frac{\langle H \Delta N \rangle}
                                                        {\langle (\Delta N)^2\rangle}.
\label{h0h1}
\end{equation}
 In this case to minimize the approximate projected energy, we have to solve the HFB equations 
with the constraint $\langle\hat{N}\rangle=N$, i.e. the old BCS recipe.
This approximation is expected to be good for very strong violating 
wave functions. In atomic nuclei where only a few pairs participate in the
pairing correlations, we do not expect the first order approximation
to be a good one.

For $M=2$ one obtains
the following equations system for the coefficients  $h_0, h_1$ and $h_2$
\begin{equation}
\left \{
\begin{array}{llll}
\langle\hat{H}\rangle & =\; h_{0}    &\; + \; 0 & + \;h_{2}\langle(\Delta\hat{N})^{2}\rangle\\ 
\langle\hat{H}\Delta{\hat{N}}\rangle & =\; 0    &\; +
\;h_{1}\langle(\Delta\hat{N})^{2}\rangle
  & \;+ \;h_{2}\langle(\Delta\hat{N})^{3}\rangle \\
\langle\hat{H}(\Delta{\hat{N}})^{2}\rangle & =\;h_{0}\langle(\Delta\hat{N})^{2}\rangle
& \;+ \;h_{1}\langle(\Delta\hat{N})^{3}\rangle
           & \;+ \;h_{2}\langle(\Delta\hat{N})^{4}\rangle 
\end{array}
\right.
\label{meq2}
\end{equation}
the solution of this equations system provides
\begin{eqnarray}
  h_{0} & = & \langle\hat{H}\rangle - h_{2} \langle(\Delta\hat{N})^{2}\rangle \\
  h_{1} & = & \frac{\langle\hat{H}\Delta\hat{N}\rangle - h_{2} \;\langle(\Delta\hat{N})^{3}\rangle}
{\langle(\Delta\hat{N})^{2}\rangle}  \label{h_1}\\
  h_{2} & = & \frac{\langle(\hat{H}-\langle\hat{H}\rangle)(\Delta\hat{N})^{2}\rangle  - 
\langle\hat{H}\Delta\hat{N}\rangle\langle(\Delta\hat{N})^{3}\rangle/\langle(\Delta\hat{N})^{2}\rangle}
{\langle(\Delta\hat{N})^{4}\rangle - \langle(\Delta\hat{N})^{2}\rangle^{2} 
- \langle(\Delta\hat{N})^{3}\rangle^{2}/\langle(\Delta\hat{N})^{2}\rangle}
\label{h_2}
\end{eqnarray}
The projected energy to second order is given by
\begin{equation}
E^{(2)}_{proj}\;=\;\langle\hat{H}\rangle\;\;-\;\;h_{2}\langle(\Delta\hat{N})^{2}\rangle
\;\;+\;\; h_{1}\;(N - \langle\hat{N}\rangle) \;\;+\;\; h_{2}\; (N - \langle\hat{N}\rangle)^{2}
\label{energy2}
\end{equation}
In  a full variation after projection method one  should vary $E^{(2)}_{proj}$ and compute
cumbersome expressions like $ \frac{\delta h_2}{\delta \Phi}$.
 In the Lipkin-Nogami prescription, however, the coefficient $h_2$ is held constant during
the variation.  As a result the variational equation is much simpler, one gets
\begin{equation}
\frac{\delta}{\delta \Phi}\langle \hat{H} \rangle - h_1 \frac{\delta}{\delta \Phi}\langle 
\hat{N}\rangle +
(N - \langle\hat{N}\rangle)\frac{\delta h_1}{\delta \Phi} -2 h_2 (N - 
\langle\hat{N}\rangle)\frac{\delta}{\delta \Phi}\langle\hat{N}\rangle -
 h_2 \frac{\delta}{\delta \Phi} \langle(\Delta\hat{N})^{2}\rangle=0.
\end{equation}
 It is obvious that the solution to this equation is given by
\begin{equation}
\frac{\delta}{\delta \Phi}\langle \hat{H} - h_2 (\Delta\hat{N})^{2}\rangle
- \lambda \frac{\delta}{\delta \Phi}\langle \hat{N}\rangle = 0,
\label{cran}
\end{equation}
 with  $\lambda$ determined by  the constraint
\begin{equation}
 \langle \hat{N} \rangle = N,
\label{concran}    
\end{equation}
{\it provided} the condition $\lambda =  h_1$ is accomplished.  This can be easily 
realized  noticing
that eq.~(\ref{cran}) must hold for any variation $|\delta \Phi \rangle$, in particular, we
can choose $|\delta \Phi \rangle = \Delta \hat{N} |\Phi \rangle$. Substitution of this specific
variation in eq.~(\ref{cran}) provides
\begin{equation}
 \langle \hat{H} \Delta \hat{N} \rangle - \lambda \langle (\Delta \hat{N})^2 \rangle -
 h_2 \langle (\Delta \hat{N})^3 \rangle =0
\end{equation}
 comparison with eq.~(\ref{h_1})  shows that $ \lambda = h_1 $.

We have formulated  the approximation  for one class of nucleons (i.e. protons 
or neutrons),  in practical cases one has to do it simultaneously for protons 
and neutrons but  it is easy to show  that for $M=2$ there is no coupling
between protons and neutrons. The energy to be minimized is
\begin{equation}
E^{(2)}_{proj} = \langle \hat{H} \rangle - h^{\rm{Z}}_2 \langle (\Delta\hat{\rm{Z}})^{2}\rangle
                                          - h^{\rm{N}}_2 \langle
					  (\Delta\hat{\rm{N}})^{2}\rangle,
\end{equation}
with the constraints $\langle \hat{\rm{Z}} \rangle = \rm{Z}$ and $\langle \hat{\rm{N}} \rangle = \rm{N}$.
In these expressions $\rm{Z}$ stands for protons and $\rm{N}$ for neutrons, $h^{\rm{Z}}_2$
is given by eq.~(\ref{h_2}), where the number operator in this equations should be
substituted by the proton number operator, $h^{\rm{N}}_2$ is defined correspondingly.  
In the same manner
if there are additional constraints, for example the angular momentum, one just
has to substitute $\hat{H}$ by $\hat{H^\prime}=\hat{H}- \omega \hat{J_x}$ 
in all equations above with $\omega$ chosen in such a way that the constraint
$\langle \hat{J_x} \rangle = [I(I+1)-\langle \hat{J_z}^2 \rangle ]^{1/2} $ 
is accomplished.

\subsection{First and second order approximation to exact particle number
projection: Derivation for density dependent forces}

Now we would like to generalize the formulae of the former section to density
dependent forces, like the Gogny force \cite{GOG.75,DEG.80,BGG.84}
\begin{eqnarray}
 v_{12}(\rho)  & = &  \sum_{i=1,2}  e^{-(\vec{r}_1-\vec{r}_2)^2/\mu_i}
\left( W_i + B_i P_\sigma -H_i P_\tau -M_i P_{\sigma} P_{\tau} \right)
 \nonumber \\
& + & i W_{LS} (\vec{\nabla}_{12} \delta (\vec{r}_1-\vec{r}_2) \wedge
\vec{\nabla}_{12}) (\vec{\sigma}_1+\vec{\sigma}_2) \nonumber \\
 & + & t_3(1+P_\sigma x_0) \delta(\vec{r}_1-\vec{r}_2)
\rho^{1/3}(\frac{\vec{r}_1+\vec{r}_2}{2}),
\label{gognyforce}
\end{eqnarray}
where the dependence on the wave function $|\Phi\rangle$, through the density 
$\rho(\vec{r})= \langle \Phi | c^\dagger (\vec{r})c(\vec{r}) |\Phi\rangle$, adds some peculiarities. 

Density dependent forces do not exhibit, in general, the same properties as the bare
two-nucleon force. Let $|\Phi\rangle$ be a $\hat{S}$-symmetry breaking
wave function and $ v_{12}(\rho)$ the corresponding interaction.
It is obvious that 
\begin{equation}
\hat{S}v_{12}(\rho) \hat{S}^{-1} = v_{12} (\hat{S}\rho \hat{S}^{-1}).
\label{tbs}
\end{equation}
That means, the two-body interaction in a $\hat{S}$-rotated system is the same
as the interaction calculated with the rotated density.
 Eq.~(\ref{tbs}) is valid for any symmetry $\hat{S}$, in the particular case
of  symmetries with no dependence on the spatial coordinate $\vec{r}$,  for
example the particle number, it also holds  
\begin{equation}
\hat{S} v_{12}(\rho)\hat{S}^{-1} = v_{12} (\hat{S}\rho \hat{S}^{-1}) \equiv
v_{12}(\rho),
\label{tbsp}
\end{equation}
i.e. the  $v_{12}(\rho)$ interaction and the generators of the symmetry $\hat{S}$
do commute. In particular, for the particle number, in which we are interested
here, $v_{12}(\rho)$ satisfies the same properties as the bare force.
 
In the context of plain mean field theories there is no ambiguity left concerning the
density dependence of the hamiltonian.
 In theories beyond mean field,  for example the Generator Coordinate Method or 
 projected theories,
in  evaluating the expectation value of the energy,  appear matrix elements
like $\langle\Phi|\hat{H}|\Phi^\prime \rangle$, or in the case of PNP,
the overlap $\langle\Phi|\hat{H} e^{i\phi \hat{N}}| \Phi\rangle$. Here 
 a complication  arises from the density dependent term of the hamiltonian
because it is not obvious which density dependence should be taken in the
potential (\ref{gognyforce}). A Skyrme force with a {\it linear} 
dependence in the density can be cast
as a three body force. For this particular case it can be shown that the density
 to be used in the hamiltonian 
 in the evaluation of the matrix elements $\langle\Phi|\hat{H}|\Phi^\prime \rangle$
must be a non-diagonal density $\rho_{ND}(\vec{r})$ given by
\begin{equation}
\rho_{ND} (\vec{r}) = \frac{\langle\Phi | c^\dagger (\vec{r}) c (\vec{r})|\Phi^\prime \rangle}
{\langle\Phi|\Phi^\prime \rangle} =  \frac{\langle\Phi |\hat{\rho}(\vec{r})|\Phi^\prime \rangle}
{\langle\Phi|\Phi^\prime \rangle},
 \label{rhoND}
\end{equation}
where we have introduced the density operator $\hat{\rho}(\vec{r})= c^+(\vec{r}) c(\vec{r})$.
 For the particular case of PNP, this density  is given by $\rho_{ND} (\vec{r}) =
 \langle\Phi | \hat{\rho}(\vec{r})
 e^{i\phi\hat{N}} |\Phi \rangle /\langle\Phi| e^{i\phi\hat{N}}|\Phi \rangle.$
 Up to now, based on this result, the same prescription has been adopted for
Skyrme forces with a {\it non linear} dependence in $\rho$
\cite{BDF.90} as well as with the Gogny force \cite{VER.96}. This prescription is,
however, not well founded. 
In the context of the LN approximation there have been two approaches to this problem.
The first one \cite{VER.96}, considers the energy as a functional of the densities (normal and
abnormal) while the second one \cite{VER.97}, is based on a many-body density dependence of
the potential. These two methods are discussed in the following subsections.

\subsubsection{Functionals of the density}

In this section we shall derive the approximate particle number projection 
equations for potentials with a density dependence like
eq.~(\ref{gognyforce}), using the prescription of appendix B
to calculate the non-diagonal matrix elements of the hamiltonian. There the use
of a density like (\ref{rhoND}) in the evaluation of non-diagonal matrix elements
of any density dependent Hamiltonian was motivated.
  
The density dependence of the hamiltonian does not modify the projected energy of
eq.~(\ref{energyM}) but only  the coefficients $h_m$ of the expansion. The non-diagonal
density causes a dependence on $\phi$ of the hamiltonian and eq.~(\ref{hphi}) looks like

\begin{equation}
 \langle\Phi|\hat{H}(\phi) e^{i\phi\hat{N}}|\Phi\rangle
 = \sum_{m=0}^{M}h_{m}\hat{\cal{N}}^m \langle\Phi|e^{i\phi\hat{N}}|\Phi\rangle,
  \label{hpn}
\end{equation}
 application of the operators $1, \hat{\cal{N}}, ..., \hat{\cal{N}}^M$ on eq.~(\ref{hpn}),
 in the limit $\phi \rightarrow 0$,  now provides
\begin{equation}
\sum_{l=0}^{n}
\left (
\begin{array}{cc}
n\\l\\ 
\end{array} \right) \left. \left\langle
(\Delta\hat{N})^l \;\frac{1}{i^{n-l}}\frac{\partial^{n-l}}{\partial\phi^{n-l}}
\hat{H}(\phi) \right|_{\phi=0} \right\rangle =\sum_{m=0}^{M}h_{m}\langle(\Delta\hat{N})^{m+n}\rangle.
\end{equation}
For   $M=1$ the energy is given by eq.~(\ref{energy1}) and the coefficients $h_0$
and $h_1$ by
\begin{eqnarray}
\langle\hat{H}\rangle &=& h_{0}   \nonumber \\
\left.\left\langle \frac{1}{i}\frac{\partial\hat{H}}{\partial\phi}\right|_{\phi=0} +
 \hat{H}\Delta\hat{N}\right\rangle & = &
     h_1\langle(\Delta\hat{N})^2\rangle,
\label{h0h1d}
\end{eqnarray}
the derivatives with respect to $\phi$ are calculated very easily, since
\begin{equation}
{\left. \frac{1}{i}\frac{\partial\rho}{\partial\phi}
\right| }_{\phi=0} = {\left.\frac{1}{i}
\frac{\partial}{\partial\phi}\frac{\langle \Phi| \hat{\rho}(\vec{r})  e^{i\phi\hat{N}}
|\Phi\rangle }{\langle \Phi|e^{i\phi\hat{N}}|\Phi\rangle } \right| }_{\phi=0}   = 
\;\langle \hat{\rho}(\vec{r})   \Delta\hat{N}\rangle 
\end{equation}
we  obtain
\begin{eqnarray}
{\left.  \frac{1}{i} \frac{\partial\hat{H}}{\partial\phi}\right|}_{\phi=0}=
  \frac{1}{i} \frac{\partial\hat{H}}{\partial\rho}{\left.\frac{\partial\rho}
{\partial\phi}\right|}_{\phi=0}=
\;\langle\hat{\rho}\Delta\hat{N}\rangle \frac{\partial\hat{H}}{\partial\rho}.
\end{eqnarray}

Substitution in eq.~(\ref{h0h1d}), provides the coefficient $h_1$
\begin{equation}
h_1 = \frac{\langle \Phi | \hat{H} \Delta \hat{N} |\Phi \rangle}
           {\langle \Phi | ( \Delta \hat{N})^2 |\Phi \rangle} +
	   \left\langle \Phi \left| 
	   \frac{\partial H(\rho)}{\partial \rho(r)}
      \frac{\langle \Phi | \hat{\rho} \Delta \hat{N} |\Phi \rangle}
           {\langle \Phi | ( \Delta \hat{N})^2 |\Phi \rangle} 
	   \right| \Phi \right\rangle.
\label{h1fd}	   	   
\end{equation}
This expression for $h_1$ is exactly the same obtained in the selfconsistent
HFB approximation, see eq.~(\ref{h1hfb}). It is interesting to notice that if
we had used the density $\langle \Phi | \hat{\rho}(\vec{r}| \Phi \rangle$, instead 
of eq.~(\ref{rhoND}), in the hamiltonian, the coefficient $h_1$ would be given by
the first term of  eq.~(\ref{h1fd}), i.e. the rearrangement term would be missing, contrary to the
 HFB result, see appendix A.

For the  $M=2$ case, we obtain
\begin{eqnarray}
\langle\hat{H}\rangle &=& h_{0}\;\;+\;\;h_{2}\;\langle(\Delta\hat{N})^{2}\rangle \nonumber \\
\left.\left\langle \frac{1}{i}\frac{\partial\hat{H}}{\partial\phi}\right|_{\phi=0} +
 \hat{H}\Delta\hat{N}\right\rangle & = &
     h_1\langle(\Delta\hat{N})^2\rangle + h_2\langle (\Delta\hat{N})^3\rangle  \label{meq4fin} \\
\left\langle \left.\left. \frac{1}{i^2}\frac{\partial^{2}\hat{H}}{\partial\phi^2}
\right|_{\phi=0} + 2\Delta\hat{N}
\frac{1}{i}\frac{\partial\hat{H}}{\partial\phi}\right|_{\phi=0}
\; +\; \hat{H}(\Delta\hat{N})^{2}\right\rangle & = & h_{0}\langle(\Delta\hat{N})^{2}\rangle +
     h_1\langle(\Delta\hat{N})^3\rangle
     + h_2\langle(\Delta\hat{N})^4\rangle. \nonumber
\end{eqnarray}
The second derivative with respect to $\phi$ is given by 
\begin{equation}
{\left.\frac{1}{i^2}\frac{\partial^2\rho}{\partial\phi^2} \right| }_{\phi=0} =
  \;\langle \Delta\hat{\rho}(\Delta\hat{N})^2\rangle 
\end{equation}
and
\begin{equation}
{\left.\frac{1}{i^2}\frac{\partial^{2}\hat{H}}{\partial\phi^{2}}\right|}_{\phi=0} =
  \langle\Delta\hat{\rho} \;\;\Delta\hat{N}\rangle^{2} 
\frac{\partial^{2}\hat{H}}{\partial\rho^{2}} + 
\langle\Delta\hat{\rho}\;\;(\Delta \hat{N})^2\rangle  \frac{\partial\hat{H}}{\partial\rho}
\end{equation}
From now on we proceed as in the non-density dependent case, i.e. we have to
solve eq.~(~\ref{cran}~) with  the constraint (\ref{concran}) but with
the coefficient $h_2$ given by the solution of equations system (\ref{meq4fin}).
It can be easily checked that the condition $\lambda = h_1$ also holds in
this case.

\subsubsection{Many-body density dependence of the Hamiltonian}

In the former section we have interpreted the Gogny potential as a functional of 
the density and we have derived, under this assumption, the approximate particle 
number projected equations. 

 Another approach \cite{VER.97} to the same problem 
is to look at the expectation value 
$\langle \Psi_N | \hat{H} |\Psi_N \rangle$ of eq.~(\ref{bigeq}) independently 
of the  fact that $|\Psi_N \rangle$ is a many body wave function. In this case one could think of  using the density 
$\langle \Psi_N|\hat{\rho}|\Psi_N \rangle/\langle \Psi_N|\Psi_N \rangle$ 
in the hamiltonian. This density, which we shall refer to as projected density 
is given by
\begin{equation}
\rho_{proj}(\vec{r})= \frac{\langle\Phi|\hat{P}^{N}\hat{\rho}(\vec{r})  \hat{P}^{N}|
\Phi\rangle }{\langle\Phi|P^{N}|\Phi\rangle}=
\frac{\langle\Phi| \hat{\rho}(\vec{r}) \hat{P}^{N}|
\Phi\rangle }{\langle\Phi|P^{N}|\Phi\rangle},
\label{rhopro}
\end{equation}
and should be used in the Gogny force (\ref{gognyforce})\footnote{
This prescription, which, as we shall see works for the particle number, may not 
work for symmetries depending on $\vec{r}$. This has to do with the fact that the particle number symmetry
is a special one because the spatial density is invariant against rotations in the gauge space
associated with $\hat{N}$, see eq.~(\ref{tbsp}).}
 in the calculation of the projected energy. We shall call $\hat{H}_I (\equiv
 \hat{H}(\rho_{proj}))$  the hamiltonian obtained with a
 dependence in the projected density. 
We would like to remark that the wave function $|\Phi \rangle$ in 
eq.~(\ref{rhopro}) is a symmetry breaking one, i.e. it is not an eigenstate of 
$\hat{N}$. This guarantees the use of generalized Hamiltonians which are
functionals of wave functions in the very large Hilbert space generated
by symmetry breaking wave functions. 
Concerning the projection, we are now able to calculate the projected energy of 
eq.~(\ref{bigeq})  and we can proceed along the same lines as in the derivation
for non-density dependent forces, i.e. to use the Kamlah expansion up to order 
$M$. All expressions are the same, one just has to substitute $\hat{H}$ by $\hat{H}_I$
 in all equations.
 
 Calculation of the projected density may be quite involved. A way out is to 
approximate the projected density to be used in the hamiltonian. This can be done in the
following way.
Substitution of (\ref{projector}) in $\rho_{proj}(\vec{r})$ provides
\begin{eqnarray}
\rho_{proj}(\vec{r})= \frac{
   \int_{0}^{2\pi}d\phi e^{-i\phi N}\langle\Phi|\hat{\rho}(\vec{r})e^{i\phi\hat{N}}
   |\Phi\rangle}
   {\int_{0}^{2\pi}d\phi e^{-i\phi N}\langle\Phi|e^{i\phi\hat{N}} |\Phi\rangle} =
    \frac{ \int_{0}^{2\pi}d\phi e^{-i\phi N}\rho(r, \phi)}
   {\int_{0}^{2\pi}d\phi e^{-i\phi N}n(\phi)},
 \label{rhopro1}
\end{eqnarray}
where we have introduced the overlap $
  \rho (\vec{r}, \phi)=\langle\Phi|\hat{\rho}(\vec{r}) e^{i\phi\hat{N}}|\Phi\rangle $.
As with the Hamilton overlap,  we may use the Kamlah expansion to approximate
the projected  density, i.e.
\begin{equation}
\rho (\vec{r}, \phi) = \sum_{m=0,M} R_{m}(\vec{r}) \hat{\cal{N}}^{m} n(\phi).
\label{rphi}
\end{equation}
Substitution of  eq.~(\ref{rphi}) in eq.~(\ref{rhopro1}) provides the projected 
nuclear density to order $M$
\begin{equation}
\rho^{(M)}_{proj}(\vec{r})=\sum_{m=0}^{M}R_{m}(\vec{r})(N-\langle\hat{N}\rangle)^{m}.
\end{equation}
The expansion coefficients $R_m$ are determined  applying the operators
$1, \hat{\cal{N}},..., \hat{\cal{N}}^M$ on 
eq.~(\ref{rphi}). In the limit $\phi \rightarrow 0$, one obtains 
\begin{equation}
\langle\Phi|\hat{\rho}(\vec{r})(\Delta{\hat{N}})^{n}|\Phi\rangle=
\sum_{m=0}^{M}R_{m}(\vec{r})\langle\Phi|(\Delta{\hat{N}})^{m+n}|\Phi\rangle,
\label{coefeqp}
\end{equation}
with $n=0,\dots,M$. To first order
\begin{equation}
\rho^{(1)}_{proj}(\vec{r}) = R_0(\vec{r}) + R_1(\vec{r})(N-\langle \hat{N} \rangle),
\end{equation}
with  $R_0= \langle \Phi |\hat{\rho}(\vec{r}) |\Phi \rangle \equiv \rho(\vec{r})$, the
HFB density, and
\begin{equation}
R_1(\vec{r}) = \frac{\langle \hat{\rho}(\vec{r})\Delta\hat{N}\rangle} 
                    { \langle (\Delta\hat{N})^2 \rangle}.
\end{equation}
Assuming that $R_1(\vec{r})(N- \langle \hat{N} \rangle)$ is a small quantity as
compared to $\rho(\vec{r})$ we can expand the density which appears in 
the Gogny force
\begin{eqnarray}
[ \rho^{(1)}_{proj}(r)]^{1/3} & & \approx \rho(\vec{r})^{1/3} 
\left[ 1 + \frac{R_1(\vec{r})}{\rho(\vec{r})}(N-\langle \hat{N} \rangle)
\right]^{1/3} \nonumber \\ 
& \approx & \rho(\vec{r})^{1/3} +\frac{1}{3}\rho(\vec{r})^{-2/3}
R_1(\vec{r})(N-\langle \hat{N} \rangle),
\label{rhoexp1}
\end{eqnarray}
substitution of this expression in the Gogny hamiltonian with the projected
density dependence, what we have called $\hat{H}_I$, provides
\begin{equation}
\hat{H}_I \equiv \hat{H}(\rho_{proj}) \approx \hat{H}(\rho)
+ \frac{\partial \hat{H}}{\partial \rho} R_1(\vec{r}) (N-\langle \hat{N} \rangle),
\end{equation}
with $\hat{H}(\rho)$ the hamiltonian depending on the HFB density.
The projected energy is given, up to $(N-\langle \hat{N} \rangle)$
order, by
\begin{equation}
E^{(1)}_{proj} = \langle \hat{H}_I\rangle  + \frac{\langle H\Delta N\rangle}
{\langle (\Delta N)^2 \rangle} (N-\langle \hat{N}\rangle)
= \langle \hat{H} \rangle + h_1^{\rm eff}(N-\langle \hat{N} \rangle)
\end{equation}
with $h_1^{\rm eff} = \frac{\langle H\Delta N\rangle}
{\langle (\Delta N)^2 \rangle} + 
\langle \frac{\partial H}{\partial \rho} R_1(\vec{r}) \rangle$.
Minimization of this energy is equivalent to
the minimization of $\langle \hat{H} \rangle - \lambda \langle \hat{N}
\rangle $ subject to the constraint $\langle \hat{N} \rangle =N$ and
$\lambda$ given by $h_1^{eff}$. Now $\lambda$ (or $h_1^{eff}$) includes
the rearrangement term present in the HFB theory with density dependent
forces. It is worth to point out that in this approach we obtain the same 
results as in the
previous section , see eq.~(\ref{h1fd}), and in the selfconsistent HFB
approximation, see eq.~(\ref{h1hfb}).

To calculate the energy to second order  one should also expand the density 
overlap to the same order. We obtain 
\begin{equation}
\rho^{(2)}_{proj}(\vec{r}) = \rho(\vec{r}) + R_{1}(\vec{r}) (N
-\langle \hat{N} \rangle) + R_{2}(\vec{r}) (N
-\langle \hat{N} \rangle)^{2} - R_{2}(\vec{r}) \langle
(\Delta\hat{N})^{2} \rangle 
\end{equation}
with    
\begin{eqnarray}
  R_{1}(\vec{r}) & = & \frac{\langle\hat{\rho}(\vec{r})\Delta\hat{N}\rangle -
  R_{2}(\vec{r}) \;\langle(\Delta\hat{N})^{3}\rangle}
{\langle(\Delta\hat{N})^{2}\rangle}  \\
  R_{2}(\vec{r}) & = & \frac{\langle(\hat{\rho}(\vec{r})-\langle\hat{\rho}(\vec{r})\rangle)(\Delta\hat{N})^{2}\rangle  - 
\langle\hat{\rho}(\vec{r})\Delta\hat{N}\rangle\langle(\Delta\hat{N})^{3}\rangle/\langle(\Delta\hat{N})^{2}\rangle}
{\langle(\Delta\hat{N})^{4}\rangle - \langle(\Delta\hat{N})^{2}\rangle^{2} 
- \langle(\Delta\hat{N})^{3}\rangle^{2}/\langle(\Delta\hat{N})^{2}\rangle}
\end{eqnarray}

As in eq.~(\ref{rhoexp1}), we can make an expansion of the dependence of the
hamiltonian on the projected density. After substitution of this expansion in
the projected hamiltonian, we obtain
\begin{equation}
\hat{H}_{I} \approx  \hat{H}(\rho ) +
\frac{\partial\hat{H}(\rho)}{\partial\rho(r)}
\left[  R_{1}(\vec{r}) (N-\langle \hat{N} \rangle) +
 R_{2}(\vec{r}) (N-\langle \hat{N} \rangle)^{2} - R_{2}(\vec{r}) \langle
(\Delta\hat{N})^{2} \rangle \right]
\label{hampro}       
\end{equation}

 The projected energy for  $M=2$ is given by 
\begin{equation}
E^{(2)}_{proj}\;=\;\langle\hat{H}_I
\rangle\;\;-\;\;{h}_{2}^{I}\langle(\Delta\hat{N})^{2}\rangle
\;\;+\;\; {h}_{1}^{I}\;(N - \langle\hat{N}\rangle) 
\;\;+\;\; {h}_{2}^{I}\; (N - \langle\hat{N}\rangle)^{2}
\label{e2pt}
\end{equation}
with 
\begin{eqnarray}
  h_{1}^{I} & = & \frac{\langle\hat{H}_{I}\Delta\hat{N}\rangle - h_{2}^{I} \;\langle(\Delta\hat{N})^{3}\rangle}
{\langle(\Delta\hat{N})^{2}\rangle}  \label{h1p}\\
  h_{2}^{I} & = & \frac{\langle(\hat{H}_{I}-\langle\hat{H}_{I}\rangle)(\Delta\hat{N})^{2}\rangle  - 
\langle\hat{H}_{I}\Delta\hat{N}\rangle\langle(\Delta\hat{N})^{3}\rangle/\langle(\Delta\hat{N})^{2}\rangle}
{\langle(\Delta\hat{N})^{4}\rangle - \langle(\Delta\hat{N})^{2}\rangle^{2} 
- \langle(\Delta\hat{N})^{3}\rangle^{2}/\langle(\Delta\hat{N})^{2}\rangle}. \label{h2p}
\end{eqnarray}
The structure of eq.~(\ref{e2pt}), is exactly the same as eq.~(\ref{energy2})
and so  are the coefficients $h^I_1, h^I_2$ and  $h_1, h_2$.
 From the discussion of eq.~(\ref{energy2}), it is obvious that the solution to this equation is given by
\begin{equation}
\frac{\delta}{\delta \Phi}\langle \hat{H_I} - h^I_2 (\Delta\hat{N})^{2}\rangle
- \lambda \frac{\delta}{\delta \Phi}\langle \hat{N}\rangle = 0,
\label{cranp}
\end{equation}
 with  $\lambda$ determined by  the constraint
\begin{equation}
       \langle \hat{N} \rangle = N,
\end{equation}
{\it provided} the condition $\lambda =  h^I_1$ is accomplished. Again, it can be checked that
this is the case. Taking into account the particle number constraint
 we can write
\[  
\hat{H}_{I} \approx  \hat{H}(\rho ) -
\frac{\partial\hat{H}(\rho)}{\partial\rho(r)} R_{2}(\vec{r}) \langle
(\Delta\hat{N})^{2} \rangle )
\]
Substitution of $\hat{H}_{I}$ in $h^I_2$ provides
\begin{eqnarray}
h_{2}^I &=& h_{2} -  \frac{\langle( \frac{\partial\hat{H}}{\partial\rho(\vec{r})}
R_{2}(\vec{r}) -\langle \frac{\partial\hat{H}}{\partial\rho(\vec{r})}
R_{2}(\vec{r}) \rangle)(\Delta\hat{N})^{2}\rangle  - 
\langle \frac{\partial\hat{H}}{\partial\rho(\vec{r})}
R_{2}(\vec{r})
\Delta\hat{N}\rangle\langle(\Delta\hat{N})^{3}\rangle/\langle(\Delta\hat{N})^{2}\rangle}
{\langle(\Delta\hat{N})^{4}\rangle - \langle(\Delta\hat{N})^{2}\rangle^{2} 
- \langle(\Delta\hat{N})^{3}\rangle^{2}/\langle(\Delta\hat{N})^{2}\rangle} \langle(\Delta\hat{N})^{2}\rangle
\nonumber \\
&=& h_{2} + \delta h_2 
\end{eqnarray}
with $h_2$ given by eq.~(\ref{h_2}).
 We have checked in some test cases that  $\delta h_2$ is much smaller
 than $h_2$ and therefore in the following  $\delta h_2$  is neglected.
  Assuming $\langle \hat{N} \rangle = N$, we can finally write the projected energy,
 eq.~(\ref{e2pt}), to be minimized
\begin{eqnarray}
E^{(2)}_{proj} &\approx& \langle \hat{H} \rangle - [ \langle  
\frac{\partial \hat{H}}{\partial\rho(\vec{r})} R_{2}(\vec{r}) \rangle + h_{2} ]
 \langle (\Delta\hat{N})^{2}\rangle \nonumber \\
&=& \langle \hat{H} \rangle - h_{2}^{eff} \langle(\Delta\hat{N})^{2}\rangle .
\end{eqnarray}
Where we have introduced the effective parameter $h_{2}^{eff}$
\begin{eqnarray}
&& h_{2}^{eff} = \frac{(\langle \hat{\cal{H}}  
-\langle  \hat{\cal{H}} \rangle )(\Delta\hat{N})^{2}\rangle
- \langle \hat{\cal{H}} \Delta\hat{N}\rangle\langle(\Delta\hat{N})^{3}\rangle
/\langle(\Delta\hat{N})^{2}\rangle}
{\langle(\Delta\hat{N})^{4}\rangle - \langle(\Delta\hat{N})^{2}\rangle^{2} 
- \langle(\Delta\hat{N})^{3}\rangle^{2}/\langle(\Delta\hat{N})^{2}\rangle}
\end{eqnarray}
with
\begin{equation}
\hat{\mathcal{H}} = \hat{H} +
 \sum_{ij} \langle \frac{\partial\hat{H}}{\partial\rho(\vec{r})}
f_{ij}(\vec{r}) \rangle c^+_i c_j
\label{heff}
\end{equation}
The quantities $f_{ij}(\vec{r})$ are those appearing in the second
quantization form of the density operator $\hat{\rho}=\sum_{ij}
f_{ij}(\vec{r}) c^+_i c_j$. The last term in this equation is a
consequence of the density dependence of the hamiltonian and does not
appear in the standard derivation of the Lipkin-Nogami approximation.
It resembles the usual rearrangement term appearing in the HFB theory
with density dependent forces.
This term might be very important for forces with density dependent
pairing.

\section{Results for ground bands of even-even nuclei.}
  
 In the evaluation of the LN energy the most complicated expression is
  $\left\langle \Delta \hat{H}(\Delta \hat{N})^{2}\right\rangle $, in appendix C
we derive the general expression for this term.
  The detailed expressions for the total LN energy and its gradient for the Gogny force 
can be found in ref.~\cite{VAL.96}. The equations have been solved using the Conjugate 
Gradient Method \cite{ELM.95}. 
  
We have applied the LN formalism with the Gogny force to study the superdeformed 
ground   bands of  even $A$  nuclei. 
We shall refer to this calculations as cranked-HFB-LN (CHFBLN).
In most results we shall also present the ones obtained with the plain  cranked
HFB theory (CHFB) \cite{ER.93}. We are using the standard D1S parameterization
set \cite{BGG.84,BGG.91}.  In some figures the theoretical results and the 
experimental data are displayed as a function of the angular momentum, in
spite of the fact that for some nuclei the spins have not been firmly assigned. The
experimental
spin values that appear in these figures are those quoted in the respective
works as the most probable ones.

To solve the CHFB and CHFBLN equations we have expanded the quasiparticles in a 
triaxial harmonic oscillator (HO) basis.
  The  HO configuration space was determined by the condition
\begin{equation}
\hbar \omega_x n_x + \hbar \omega_y n_y + \hbar \omega_z n_z \leq \hbar \omega_0 N_0
\end{equation}
where $n_x, n_y$ and $n_z$ are the HO quantum numbers and the frequencies
$\hbar \omega_x, \hbar \omega_y$ and  $\hbar \omega_z$ are determined by
$\omega_x = \omega_y = \omega_0 q^{\frac{1}{3}},\; \omega_z = \omega_0 q^{-\frac{2}{3}}$.
The parameter $q$ is strongly connected to the ratio between the nuclear size along  the
$z$- and  the perpendicular direction. A value of $q=1.5$ is a good election for superdeformed 
nuclei. For $ N_0$ we have taken a value of  $N_0 =12.5$, which provides a basis big enough as to warrant the
convergence of the results.

As it is usually done, to save CPU time, the following terms of the 
interaction have not been taken into account in the calculations~:
 The Fock term of the Coulomb interaction and the contributions to the 
 pairing field of the spin-orbit, Coulomb and the two-body center of 
mass correction terms.   
   
   It is important to notice that in the calculations of the coefficient
 $h_2$ of the Kamlah expansion, the full hamiltonian is used in the expressions.
Concerning the two prescriptions to deal with the density dependence of the
hamiltonian, in ref.~\cite{VER.97} it was shown that for the $ A=190$ region both 
prescriptions provide very similar results. Therefore, for this region we show
results only with the prescription of the many body density dependence of the hamiltonian.
For the $A=150$ region we shall show results for both prescriptions.

\subsection{The $A=190$ region.}

\begin{figure}[h]
\begin{center}
\parbox[c]{14cm}{\includegraphics[width=6cm,angle=270]{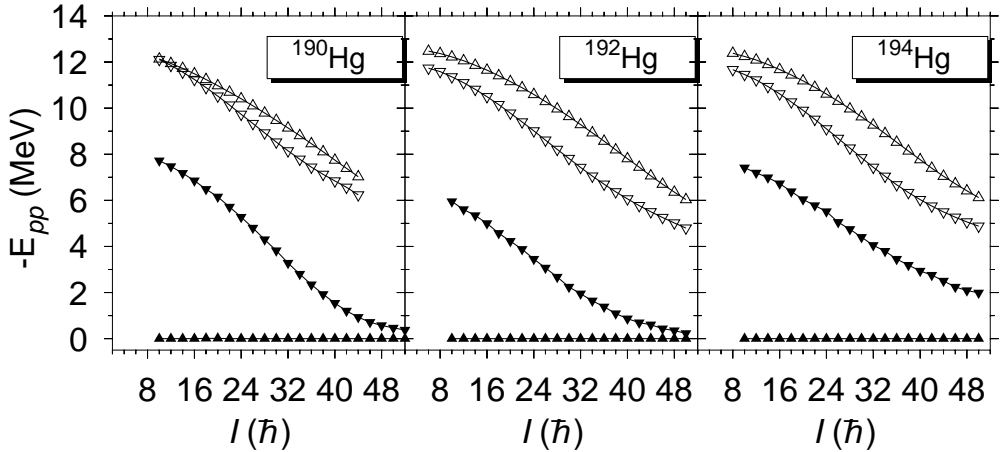}}
\end{center}
\caption{Particle-particle correlation energies for $^{190-192-194}$Hg.
 Full triangles~: CHFB  pairing energy for protons (upwards) and neutrons (downwards).
Empty triangles~: CHFBLN  pairing energy for protons (upwards) and neutrons
(downwards).}
\label{fig1hg}
\end{figure}

  In this section we shall present results for the ground band of the nuclei
 $^{190-192-194}$Hg and $^{194}$Pb. 
In fig.~\ref{fig1hg} we show the particle-particle correlation energy,
defined by $E_{pp}=\frac{1}{2} Tr(\Delta \kappa)$,
in the CHFB  and in the CHFBLN approaches, versus the angular momentum, for 
the nuclei $^{190-192-194}$Hg.

In the CHFB approach we do not find pairing correlations for the proton system
for the nuclei considered. For the neutron system we find sizable pairing
correlations at low and medium spins which diminish at large angular momentum
due to a strong Coriolis antipairing effect,  at very high spins
they get very small.
 In the CHFBLN approach, we find larger correlation energy for both, the proton
and neutron systems, we also observe the  Coriolis antipairing effect
as the angular momentum increases but the systems remain correlated even
at the largest spins. We also realize that the LN term has a larger effect on 
the proton system than on the neutron one. 

\begin{figure}[b]
\begin{center}
\parbox[c]{7cm}{\includegraphics[width=7cm]{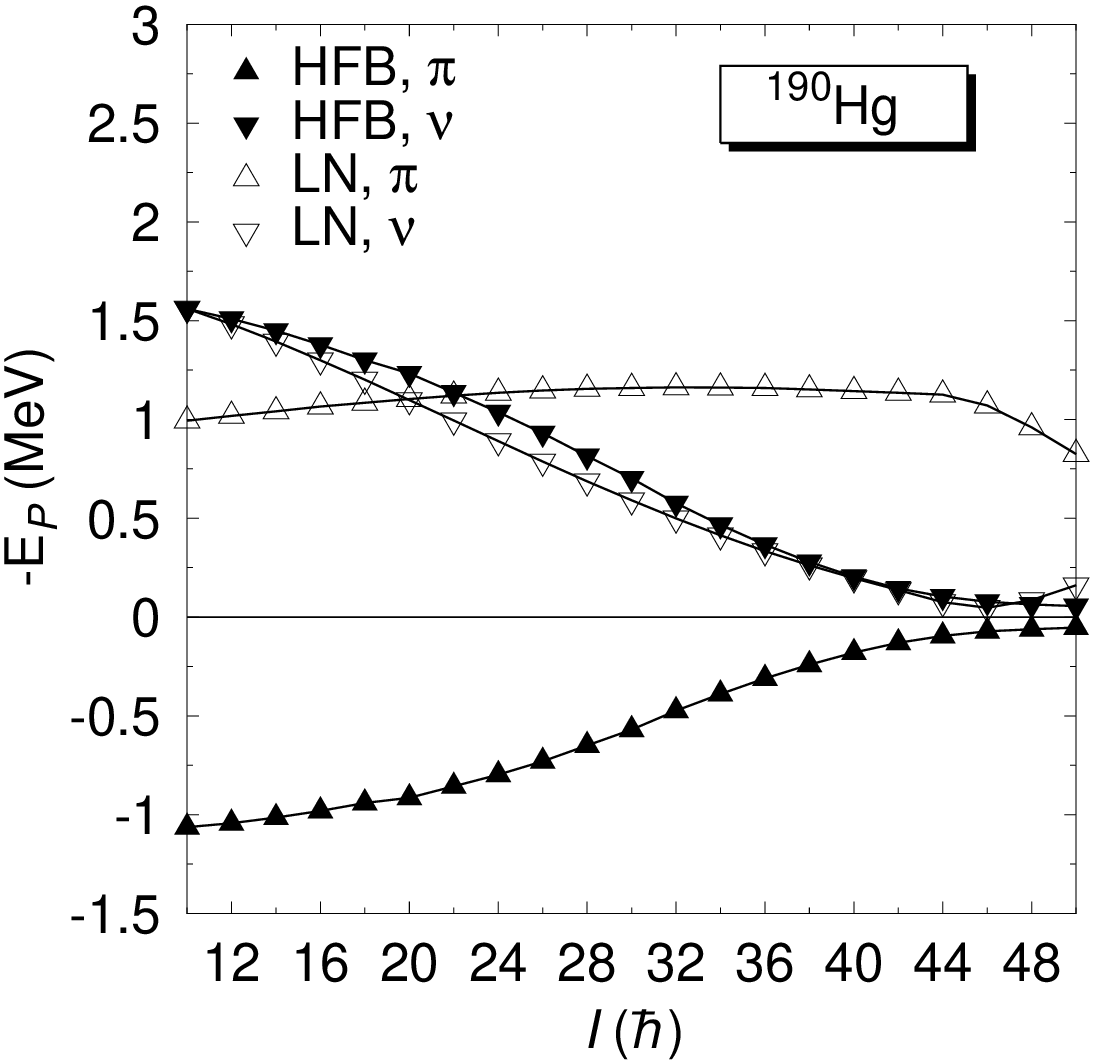}}
\end{center}
\caption{Pairing energies,  eq.~(\ref{E_P}), for $^{190}$Hg. Full triangles~: 
CHFB  pairing energy for protons (upwards) and neutrons (downwards).
Empty triangles~: CHFBLN  pairing energy for protons (upwards) and neutrons
(downwards).}
\label{fig2phg}
\end{figure}

The energy $E_{pp}$ is a
measure of the pairing correlations, but there are other ways to look for
pair correlations. In the scientific literature, and in the scope of
HFB theories, the pairing energy $E_P$ is
defined as
\begin{equation}
E_P = E - E_{HF}
\label{E_P}
\end{equation}
with $E$ the energy calculated in the HFB (or LN) approximation and $E_{HF}$
the energy calculated in the Hartree-Fock approach.
 This quantity is different from 
what we have called particle-particle correlation energy $E_{pp}$. The quantities
$E_P$ and $E_{pp}$ coincide only in the case of a BCS calculation 
performed on top of a HF solution, i.e. without selfconsistency in the 
Hartree-Fock and the pairing fields. 

\begin{figure}[h]
\begin{center}
\parbox[c]{14cm}{\includegraphics[width=10cm,angle=270]{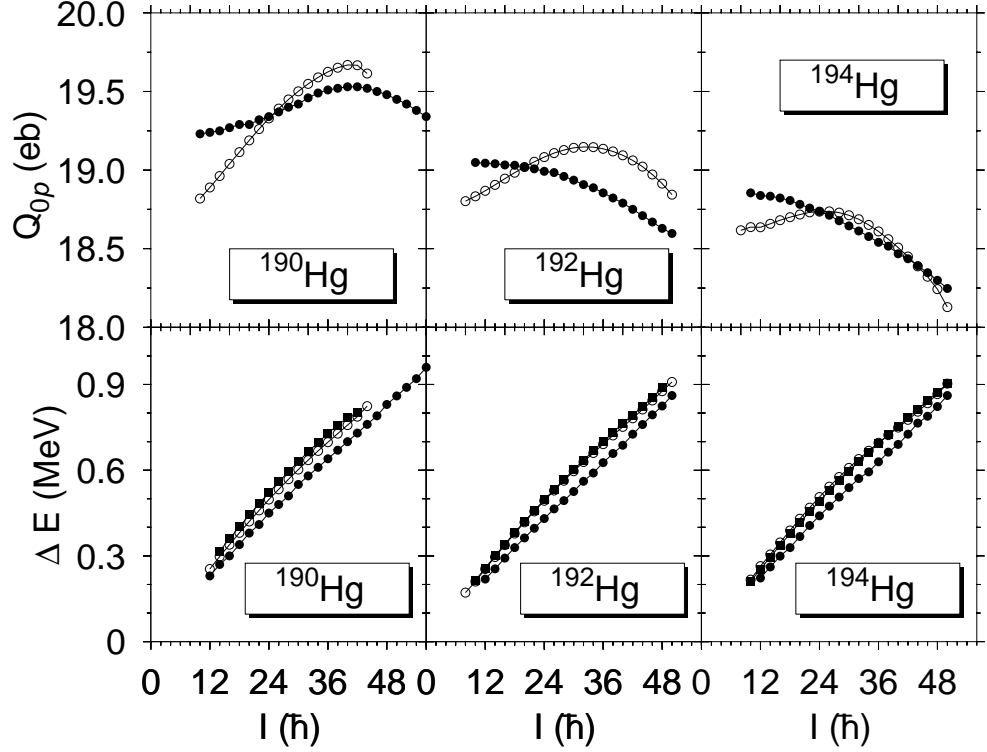}}
\end{center}
\caption{Results for $^{190-192-194}$Hg. Upper panel~: The charge
quadrupole moment versus the angular moment. Full (empty) circles
represent the CHFB (CHFBLN) approach.
Lower panel~: Gamma-ray energy
 $\protect \Delta E_I=E(I)-E(I-2)$ as a function of the angular momentum.
Full (open) circles stand for the CHFB (CHFBLN) results and full
squares  for the experimental data, for $^{190}$Hg \protect \cite{CRO.95},
for $^{190}$Hg \protect \cite{FAL.95} and for  $^{194}$Hg \protect \cite{CED.94} .}
\label{fig3hg}
\end{figure}

In fig.~\ref{fig2phg}, as an example, we display the quantity $E_P$ for $^{190}$Hg,
 in the
HFB and the LN approach for protons and neutrons. In the HFB solution we
have found a superfluid solution in the neutron channel but not in the 
proton channel. The HF solution provides the proton and neutron densities,
 $\rho_Z$ and $\rho_N$, that minimize the energy in the absence of pairing 
correlations.
We expect, therefore, the proton contribution to the total energy to be
deeper in the HF than in the HFB approach, the neutron contribution, on the
contrary, must be deeper in the HFB than on the HF approach. The total energy
must be, obviously, deeper in the HFB than in the HF approach. We observe in 
fig.~\ref{fig2phg} that these expectations are fulfilled. We also find a slow
convergence to zero in the energy $E_P$, as a function of the angular momentum,
 indicating the
quenching of the pairing correlations. In the LN approach we find the neutron
system to behave very similarly to the HFB one, the proton one behaves
differently as expected from the behavior of $E_{pp}$. It is interesting to
point out that $E_P$, a quantity with more physical content than $E_{pp}$,
is much smaller than $E_{pp}$ as one would expect from the experimental
data.  It is important to notice that whereas the correlation energy gained by 
the HFB procedure with respect to HF is only of 0.5 MeV at $I=12 \hbar$,
the correlation energy gained by the LN procedure with respect to HFB at the 
same spin is of 2.0 MeV. This result indicates the relevance of the restoration
of the particle number symmetry in this case.

\begin{figure}[h]
\begin{center}
\parbox[c]{12cm}{\includegraphics[width=10cm,angle=270]{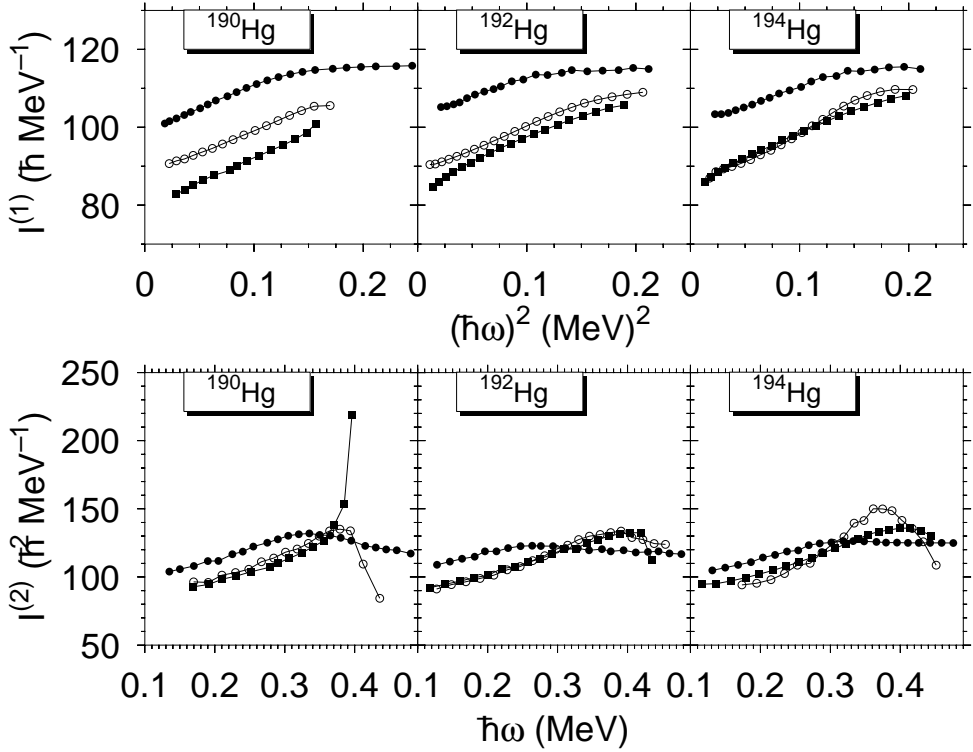}}
\end{center}
\caption{ Upper panel~: the static  moment of inertia,
$\protect {\cal I}^{(1)}(I)= (2I-1)/\Delta E_I$ versus the square of the angular frequency,
$\protect \hbar \omega (I) = \Delta E_I/2$.
Lower panel~: Second moments of inertia,
$\protect {\cal I}^{(2)}(I)= 4/(\Delta E_I- \Delta E_{I-2})$,
 for $^{190-192-194}$Hg.
 The meaning of the symbols is the following~:
Squares  for the experimental data \protect \cite{CRO.95}, \protect \cite{FAL.95},
 \protect \cite{CED.94}, full (empty) circles for the CHFB (CHFBLN) approach.}
\label{fig4hg}
\end{figure}

  In the upper part of fig.~\ref{fig3hg} we show the angular momentum
dependence of the quadrupole moment, $Q_{0p}$, for the nuclei $^{190-192-194}$Hg.
With increasing neutron number we observe a global decrease
in deformation  and a  reduction in the $I$-values in which the quadrupole 
moment starts to decrease. The CHFBLN and the CHFB results
differ mostly at low spins where the pairing correlations are larger.
This causes the crossing of the results of both calculations for each nucleus.
As a function of the angular momentum we observe first an increase of the
quadrupole moment due to a reduction of the pairing correlations and then a 
decrease of $Q_{0p}$. This anti-stretching effect is caused by the Coriolis force. 
 The  CHFB and the CHFBLN  predictions are slightly higher than the 
experimental values $17.7^{+1}_{-1.2}$ eb for$^{190}$Hg \cite{AMR.97}, 
$17.7\pm 0.8$ eb for$^{192}$Hg \cite{MOO.97} and
$17.7\pm 0.4$ eb for$^{194}$Hg \cite{MOO.97}. 
These quadrupole moments correspond to $\beta$ deformations along the yrast band
 of $0.560-0.580$,
$0.555-0.565$, $0.540-0.552$ for $^{190-192-194}$Hg respectively, the gamma
deformation remains zero for the three nuclei at all spin values.

In the lower part of fig.~\ref{fig3hg} we display the transition energies versus the
angular momentum.  
The agreement with the experimental data is good in the CHFBLN approach, in the
CHFB  one obtains too low values. The reason for this behavior is again the smallness of the
pairing correlation  in the CHFB approach as it can be seen in the upper part of 
fig.~\ref{fig4hg} where we display the static moments of inertia versus the square of
the angular frequency $\hbar \omega$. Here we observe that the CHFB moments of inertia
are too large as compared with the experiment. The CHFBLN moments of inertia are in
much better agreement with the experiment, specially for $^{192-194}$Hg. 

\begin{figure}[htb]
\begin{center}
\parbox[c]{10cm}{\includegraphics[width=10cm,angle=0]{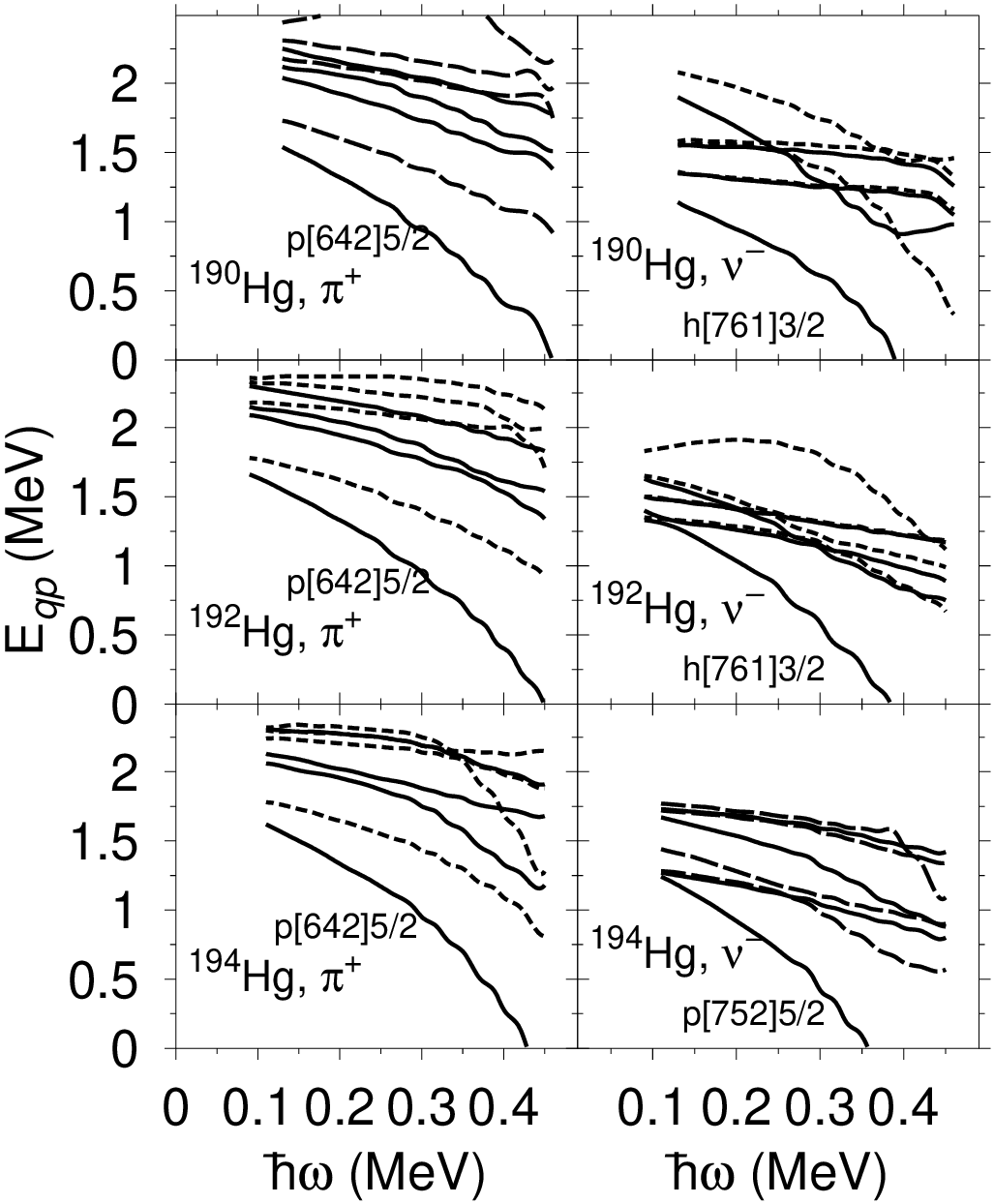}}
\end{center}
\caption{Quasiparticles energies for the nuclei  $^{190}$Hg (upper panel)
 $^{192}$Hg (middle panel) and $^{194}$Hg (lower panel), in the left hand side for
positive parity protons and in the right hand side for negative parity
neutrons. The continuous (dashed) lines
correspond to positive (negative) signature.}  
\label{fig5hg}
\end{figure}

  In the lower part of fig.~\ref{fig4hg}, the second moment of inertia
${\cal I}^{(2)}$ as a function of the angular frequency is represented.
Here, as in the previous picture, the CHFB results are in poor agreement with
the experiment. At small spin values the moment of inertia is too large and 
the neutron alignment sets in too early.
 The CHFBLN results, however, are in very good agreement with the
experimental results, the correlations induced by the Lipkin-Nogami prescription
produce the desired effect, namely, to diminish the moment of inertia at small
spin values and to delay the alignment to higher spin values. 
 For $^{190}$Hg the experimental results show a sharp upbending at $I=42 \hbar$
 (~$\hbar \omega \approx 0.4$ MeV~), in our calculations we do not get a sharp
upbending but a modest one, we see however a clear discontinuity in the ${\cal I}^{(2)}$
at $I=44\hbar$.  For the last points of our calculations there are not experimental 
results.
One expects, however, that after the sharp backbending the experimental results will
bend down. The CHFBLN for $^{192}$Hg are in very good agreement with the experimental
ones and for $^{194}$Hg we get slightly more alignment at high spins than in the experiment.

To investigate the orbitals responsible for the alignments we display in
 fig.~\ref{fig5hg} the quasiparticle energies for the three nuclei. We find that
the orbitals responsible for the alignment at $\hbar \omega \approx 0.4$ MeV are 
the $N=7$ ones, mainly  the $\nu [761]\frac{3}{2}$ for $^{190}$Hg
and  $^{192}$Hg and the $\nu [752]\frac{5}{2}$ for $^{194}$Hg, though it is
hard to distinguish both orbitals because they are strongly mixed. These
results are in good agreement with the ones obtained in \cite{THB.95}. 
The proton alignment takes place at higher
frequencies ($\hbar \omega \approx 0.45$).

\begin{figure}[htb]
\begin{center}
\parbox[c]{14cm}{\includegraphics[width=9cm,angle=270]{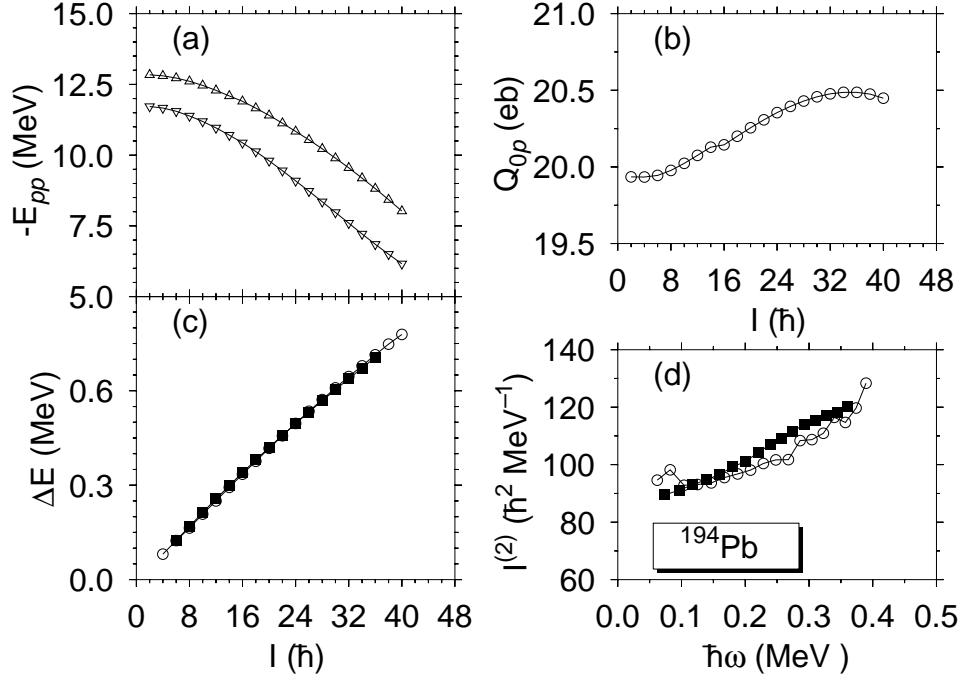}}
\end{center}
\caption{ The nucleus  $^{194}$Pb in the CHFBLN approximation. (a) The 
particle-particle correlation energy for protons (triangles) and neutrons
(inverted triangles) (b) The charge quadrupole moment,
(c) the transition energy, the experimental data are from \protect \cite{KRU.95}
 and (d) the dynamical moment of inertia}  
\label{fig6pb}
\end{figure}

As a further example of this region we display in fig.~\ref{fig6pb} the 
results for the nucleus $^{194}$Pb in the CHFBLN approximation. The particle-particle
correlation energy $E_{pp}$, fig.~\ref{fig6pb}a, shows a similar behavior to the Hg isotopes,
i.e., a modest Coriolis antipairing effect at small spins which becomes larger
at high angular momentum. The
quadrupole moment behaves similarly to the previous studied nuclei~: first, 
it increases as a consequence of the
weakening of the pairing correlations and at high angular momentum starts
to decrease due to the anti-stretching effect. The numerical value of $Q_{0p}$ is also
a little larger than the experimental value of $18.8 \pm 1.1$ eb, \cite{KRU.97}. The transition energies 
are depicted in fig.~(\ref{fig6pb}c), here the agreement with the
experimental data is excellent. Finally, in fig.~(\ref{fig6pb}d) the dynamical moment
of inertia is shown, we find a very good agreement with the experimental
data at low and high angular momentum. The  analysis of the quasiparticles energies
as a function of the angular momentum, indicates that in this case the orbital responsible for the
alignment at high spin is the $\nu [752]\frac{5}{2}$.

\begin{figure}[h]
\begin{center}
\parbox[c]{14cm}{\includegraphics[width=9cm,angle=270]{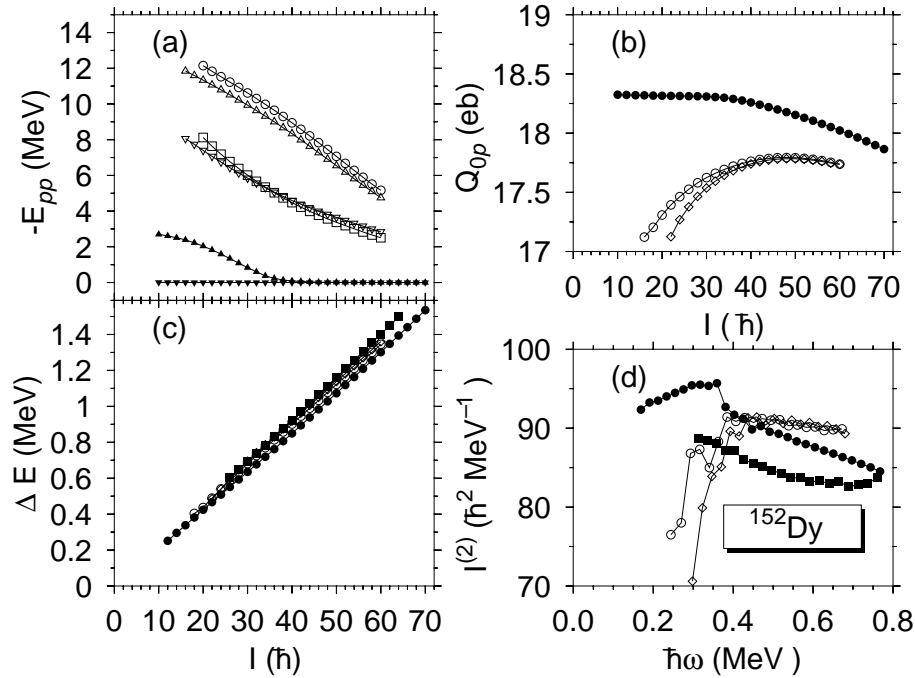}}
\end{center}
\caption{ The nucleus  $^{152}$Dy in the LN and the HFB approximations. (a)
The particle-particle correlation energy in CHFBLN (CHFBLNFD) for protons, 
open triangles
(circles) and neutrons, empty inverted triangles (squares).
 Same quantity in CHFB~: protons, triangles; neutrons, inverted triangles.
   (b) The charge quadrupole moment, CHFBLN (empty circles), CHFBLNFD (squares)
 and CHFB (full circles).
(c) the transition energy. Experiment (full squares) \protect \cite{DAG.94}, 
CHFBLN (empty circles), CHFBLNFD (empty squares),  CHFB (full circles) 
 and (d) the dynamical moment of inertia. Same symbols as in (c).}  
\label{fig7dy}
\end{figure}

\subsection{The $A=150$ region.}

 We shall now turn to the $A=150$ region. This region is characterized by 
smaller pairing correlations, the restoration of particle number symmetry should
not be, therefore, as relevant as in the $A=190$ region. As examples of
this region we discuss the nuclei $^{152}$Dy and $^{150}$Gd.

As mentioned at the beginning of this section, we would like to compare,
in the $A=150$ region,
the results obtained with the two approaches to the problem of the density dependence
of the Hamiltonian discussed in the theory section.

In fig.~\ref{fig7dy} we represent the main results for the nucleus $^{152}$Dy
in the HFB and in the LN approach. The LN results with the prescription of the
many-body density dependence we shall call CHFBLN, as before,
and those based in the  functional of the density prescription, CHFBLNFD.

\begin{figure}[h]
\begin{center}
\parbox[c]{7cm}{\includegraphics[width=7cm,angle=0]{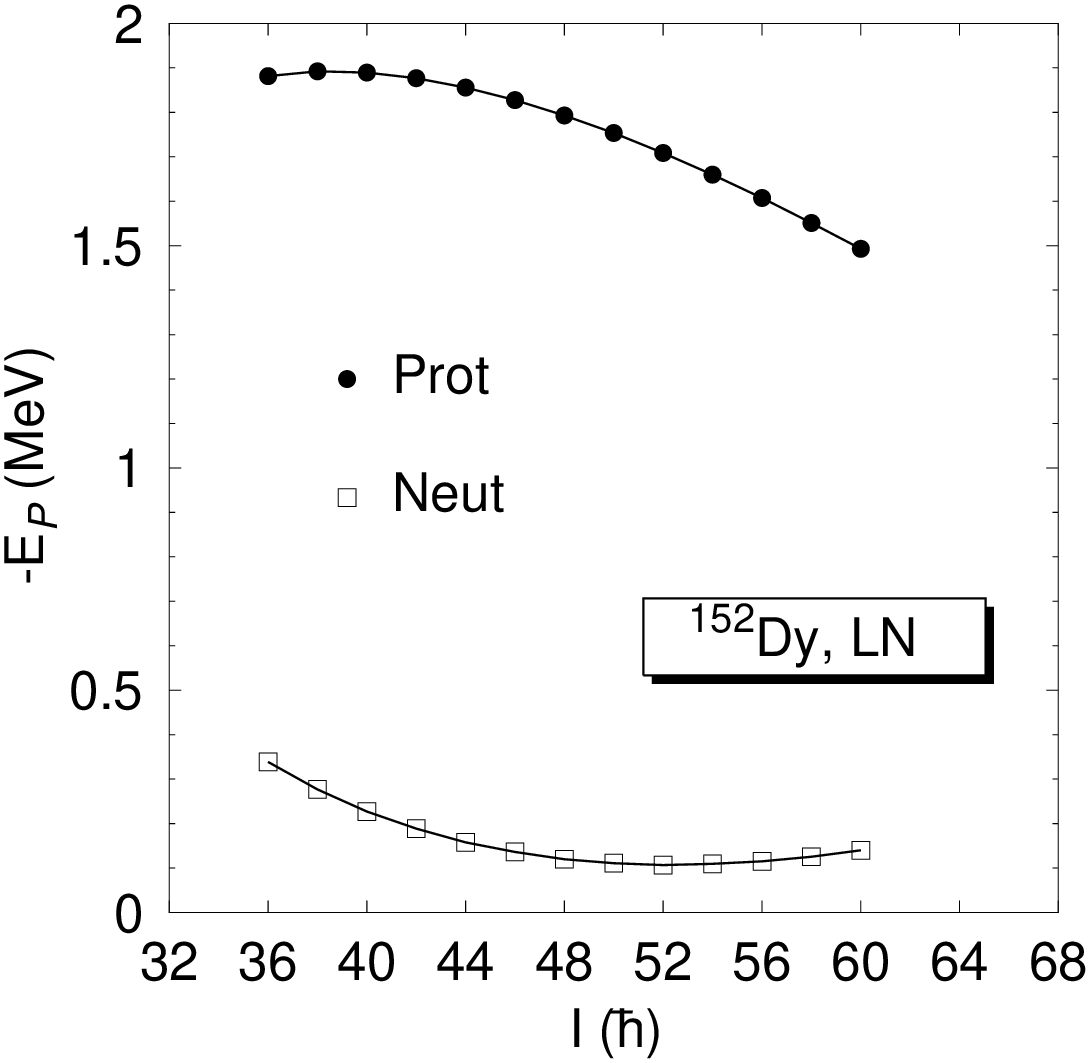}}
\end{center}
\caption{ Pairing energies,  Eq.~(\ref{E_P}), for $^{152}$Dy in the CHFBLN 
approximation for protons (dots) and neutrons (squares).} 
\label{fig8dy}
\end{figure}

The particle-particle correlation energy, $E_{pp}$, is represented in 
fig.~\ref{fig7dy}a as a function of the angular momentum. In the CHFB approximation,
the neutron system is not
correlated while the proton one is slightly correlated  for spin 
values smaller than $36 \hbar$. In the LN approximation, however both
systems are correlated. The  results in the CHFBLNFD and CHFBLN are very similar
for protons and neutrons.
As expected we obtain smaller correlation energies in this region than in the
$A=190$ one. In  fig.~\ref{fig8dy} we plot the energy $E_P$ of eq.~(\ref{E_P})
in the CHFBLN approach. For  neutrons is very small for all spin values and for
protons around 1.8 MeV at spin $40 \hbar$ and it decreases to 1.5 MeV for spin
$60\hbar$, i.e. the energy gain allowing superfluid wave functions is about 2 MeV.

In fig.~\ref{fig7dy}b we display the charge quadrupole moment, $Q_{0p}$, versus the
angular momentum. In the CHFB approximation and at small angular momentum 
it is almost constant, only around spins values of $30\hbar$, $Q_{0p}$ starts to
decrease by the Coriolis anti-stretching effect. In the CHFBLN approaches the
behavior at low spins is quite different (it increases due to the Coriolis
antipairing effect) and only around spin values of $50\hbar$ starts to
decrease. The experimental value is $17.5 \pm 0.2$ eb, \cite{SKW.96,NJM.97},
in good agreement with the  CHFBLN results, the HFB values being somewhat larger.
The nuclei are axially symmetric, i.e. the gamma deformation is zero.
In  fig.~\ref{fig7dy}c the gamma ray energies are plotted versus the angular
momentum. The HFB predictions are the lowest ones due to the large moment of
inertia. The CHFBLN results are slightly below the experimental data and almost parallel 
to them. Both CHFBLN approaches to the gamma ray energies practically coincide. 
Finally in fig.~\ref{fig7dy}d the second moments of inertia are
displayed versus the angular frequency. The CHFB results are larger than
the experimental ones but they behave in a similar way. The inclusion of
additional pairing correlations through the LN approach produces an
enhancement of the second moment of inertia as compared with the CHFB results.
This anomalous behavior has been also found in \cite{BFH.96}. The two LN approaches
are very similar at high spin values, only at small spins one finds differences.
 The low values of ${\cal I}^{(2)}$ at small angular frequencies are probably
 due to an excess of pair correlations at low $I$, as one can also see in 
 the behavior of the quadrupole moment, fig.~\ref{fig7dy}b,  ${\cal I}^{(2)}$ being
 a second derivative magnifies small effects. The first moment of inertia,
  fig.~\ref{fig7dy}c, however, has very  reasonable values.
 Concerning the quality of the LN approaches, we may conclude, as in the 
 $A=190$ region that both approaches practically coincide.

\begin{figure}[htb]
\begin{center}
\parbox[c]{10cm}{\includegraphics[width=8cm,angle=270]{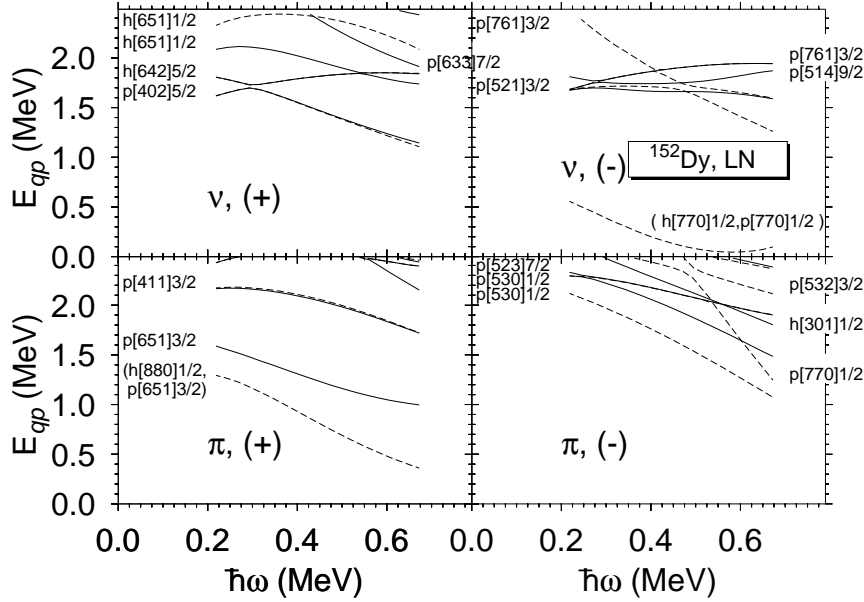}}
\end{center}
\caption{Quasiparticles energies for the nucleus  $^{152}$Dy, in the upper part for
protons and in the lower part for neutrons. The continuous (dashed) lines
correspond to positive (negative) signature.}  
\label{fig9dy}
\end{figure}

In fig.~(\ref{fig9dy}) the quasiparticle energies for the nucleus $^{152}$Dy
are represented versus de angular frequency. From these levels one can predict the 
blocking structure of the excited bands in this nucleus or in the nearby odd-even nuclei. 
One should be aware, however, that there is strong mixing within the Nilsson levels
and that the character of particle or hole  near the Fermi surface, in the presence
of pairing correlations, is not a precise statement. 
Most of the levels shown in the figure coincide with the assignment proposed 
\cite{DAG.94,AKR.96}.

\begin{figure}[h]
\begin{center}
\parbox[c]{14cm}{\includegraphics[width=9cm,angle=270]{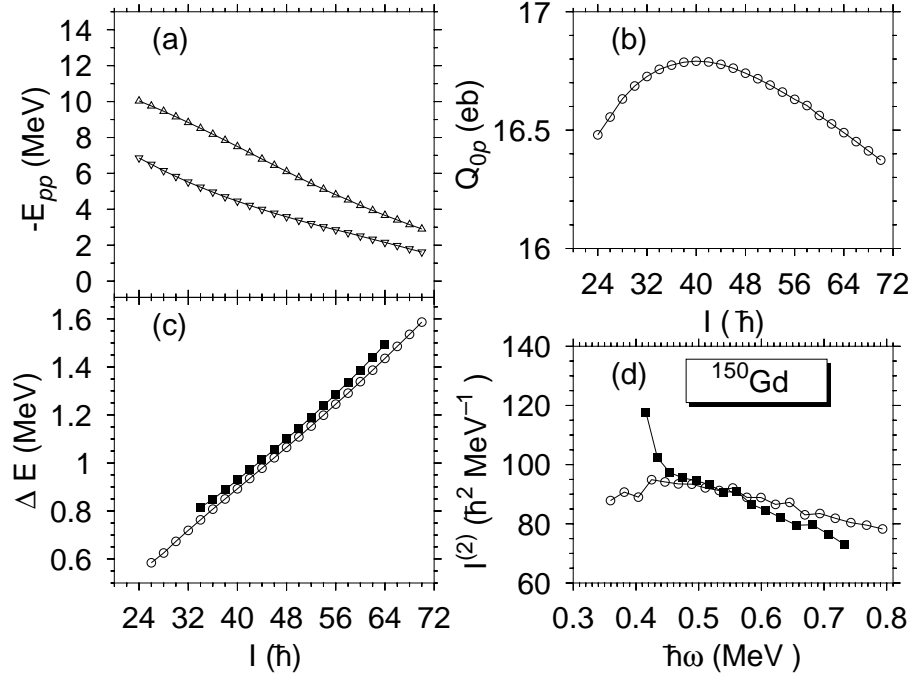}}
\end{center}
\caption{ The nucleus  $^{150}$Gd in the CHFBLN approximation.
(a) The 
particle-particle correlation energy for protons (triangles) and neutrons
(inverted triangles) (b) The charge quadrupole moment,
(c) the transition energy, the experimental data \protect \cite{FAL.89}
 are represented by squares and (d) the dynamical moment of inertia}  
\label{fig10gd}
\end{figure}

The next example in the $A=150$ region is the $^{150}$Gd, the results are shown in
fig.~\ref{fig10gd}. Since we have seen
in the previous example that both density prescriptions provide similar results,
for this nucleus we only show the results with the many body density prescription.
In fig.~\ref{fig10gd}a, the CHFBLN pairing energies versus the angular momentum
are represented, they
show an smooth decrease with growing angular momentum. In fig.~\ref{fig10gd}b,
the charge quadrupole moments as a function of $I$ are presented. 
They display the typical Coriolis effects: first an increase caused by the 
Coriolis antipairing effect and later one the Coriolis antistretching effect. 
It is in good agreement with the experimental value, $17^{+0.5}_{-0.4}$ eb, 
\cite{BFA.98}.
 This nucleus is also axially symmetric.
The transition energies represented in fig.~\ref{fig10gd}c are well reproduced
by the theoretical predictions, indicating, nevertheless, a slightly larger
moment of inertia than the experiment. The second moment of
inertia is shown in fig.~\ref{fig10gd}d, the agreement is good except at
the lowest values of the angular momentum, where the experimental points
present a sharper level crossing than in the theory.

\section{Conclusions}

In conclusion, we have discussed the Lipkin-Nogami equations for density dependent
forces and two different prescriptions for the density dependence of the Hamiltonian  
to be used in the evaluation of non-diagonal matrix elements.

As an application we have analyzed the high spin properties of the nuclei 
 $^{190-192-194}$Hg and $^{194}$Pb of the $A=190$ region and the
 nuclei $^{152}$Dy and $^{150}$Gd of the $A=150$ region in the Cranked 
 Hartree-Fock-Bogoliubov and the Cranked Lipkin-Nogami approaches. 
 
 Global properties as well  as sensitive properties as the second moment of 
inertia are  well reproduced, specially within the  CHFBLN approximation. 
In general we obtain excellent qualitative agreement with the experiment
 and in some cases a quantitative one. The approximate restoration of the particle
number symmetry is the essential ingredient for the quantitative agreement.
We believe that the Gogny force with the approximate particle
number projected mean field theory provides a good description of nuclear properties 
under a  large variety of situations.

This work was supported in part by DGICyT, Spain under Project  PB97--0023.
One of us (A.V.) would like to thank the Spanish Ministerio de Asuntos
Exteriores for financial support through an ICI grant.

\section{Appendix A~:Lagrange multiplier with density dependent forces}
   In the HFB theory the wave function $|\Phi\rangle$ that minimizes the energy
   with the constraint of providing the right particle number on the average
   is given by
\begin{equation}
 \delta \langle \Phi | \hat{H} - \lambda \hat{N}|\Phi\rangle =0
\end{equation}
for density dependent forces one obtains
\begin{equation}
\langle \delta \Phi | \hat{H}|\Phi\rangle +
 \left\langle \Phi \left| \frac{\partial H(\rho)}{\partial \rho(r)}
      \langle\delta  \Phi | \hat{\rho} |\Phi \rangle
	   \right| \Phi \right\rangle
-  \lambda \langle \delta \Phi |\hat{N}|\Phi\rangle =0
\label{delener}
\end{equation}
this equation must hold for any arbitrary variation $|\delta \Phi \rangle$, in
particular for $|\delta \Phi \rangle = \Delta \hat{N} |\Phi \rangle$,
substitution in eq.~(\ref{delener}) provides
\begin{equation}
\lambda = \frac{\langle \Phi | \hat{H} \Delta \hat{N} |\Phi \rangle}
           {\langle \Phi | ( \Delta \hat{N})^2 |\Phi \rangle} +
	   \left\langle \Phi \left| \frac{\partial H(\rho)}{\partial \rho(r)}
      \frac{\langle \Phi | \hat{\rho} \Delta \hat{N} |\Phi \rangle}
           {\langle \Phi | ( \Delta \hat{N})^2 |\Phi \rangle} 
	   \right| \Phi \right\rangle.
\label{h1hfb}	   	   
\end{equation}


\section{Appendix B~: Functionals of the density}


We want to define a prescription to evaluate overlaps of the hamiltonian 
\[
\frac{\left\langle \Phi |\hat{H}|\Phi '\right\rangle }{\left\langle \Phi |\Phi '\right\rangle },\]
for density dependent forces where both \( \left| \Phi \right\rangle  \) and
\( \left| \Phi '\right\rangle  \) are product wave functions of the HFB type.
The philosophy behind the prescription is that a density dependent force does
not define a hamiltonian but rather the energy, which is a functional of the
density and the pairing tensor 
\begin{equation}
E[\rho ,\kappa ]=\sum _{k_{1}k_{2}}T_{k_{1}k_{2}}\rho _{k_{2}k_{1}}+
\frac{1}{4}\sum _{k_{1}k_{2}k_{3}k_{4}}\overline{\nu}_{k_{1}k_{2}k_{3}k_{4}}\left( 2\rho _{k_{4}k_{2}}\rho _{k_{3}k_{1}}+\kappa _{k_{3}k_{4}}\kappa ^{*}_{k_{1}k_{2}}\right) 
\label{erk}
\end{equation}
where the two-body interaction could depend upon the spatial density \( \rho (\vec{r}) \)
to some power \( \alpha  \), \( v[\rho ^{\alpha }(\vec{r})]\),
 as is the case for the Gogny force. In the above expression the spatial density
is defined in the usual way 
\[
\rho (\vec{r})=\frac{\left\langle \Phi |\hat{\rho }|\Phi \right\rangle }{\left\langle \Phi |\Phi \right\rangle
}=\sum _{k_{1}k_{2}}f_{k_{1}k_{2}}(\vec{r})\rho _{k_{2}k_{1}},\]
where the density operator is given by
\[
\hat{\rho }(\vec{r})=\sum ^{A}_{i=1}\delta (\vec{r}-\vec{r}_{i})=\sum
_{k_{1}k_{2}}f_{k_{1}k_{2}}(\vec{r})c^{+}_{k_{1}}c_{k_{2}}.\]
 
The formal evaluation of the hamiltonian overlap for {\it non-density dependent forces} 
is straightforward. Using the extended Wick theorem, one gets
\begin{equation}
\frac{\left\langle \Phi |\hat{H}|\Phi '\right\rangle }{\left\langle \Phi |\Phi '\right\rangle }=\sum _{k_{1}k_{2}}T_{k_{1}k_{2}}\widetilde{\rho }_{k_{2}k_{1}}+
\frac{1}{4}\sum
_{k_{1}k_{2}k_{3}k_{4}}\overline{\nu}_{k_{1}k_{2}k_{3}k_{4}}\left(
2\widetilde{\rho }_{k_{4}k_{2}}\widetilde{\rho }_{k_{3}k_{1}}+\widetilde{\kappa
}_{k_{3}k_{4}}\widetilde{\overline{\kappa }}_{k_{1}k_{2}}\right),
\label{ove}
\end{equation}
where
\begin{equation}
\widetilde{\rho }_{k_{3}k_{1}}  =  \frac{\left\langle \Phi
|c^{+}_{k_{1}}c_{k_{3}}|\Phi '\right\rangle }{\left\langle \Phi |\Phi
'\right\rangle}, \;\;\;
\widetilde{\kappa }_{k_{3}k_{4}}  =  \frac{\left\langle \Phi
|c_{k_{4}}c_{k_{3}}|\Phi '\right\rangle }{\left\langle \Phi |\Phi '\right\rangle
},\;\;\;
\widetilde{\overline{\kappa }}_{k_{1}k_{2}}  =  \frac{\left\langle \Phi
|c^{+}_{k_{1}}c^{+}_{k_{2}}|\Phi '\right\rangle }{\left\langle \Phi |\Phi
'\right\rangle }.
\label{rhotilde}
\end{equation}
 Therefore, for {\it  non-density dependent forces} both, the energy, eq.~(\ref{erk}) and the hamiltonian
overlap, eq.~(\ref{ove}), are given by the {\it same functional} of the density matrix and the pairing tensor
but evaluated with the corresponding densities and pairing tensors. Inspired by these
results we  use the following prescription for {\it density dependent forces}~: in the evaluation 
of the hamiltonian overlap  a
density dependent interaction depending upon \( \widetilde{\rho } \), eq.~(\ref{rhotilde}), must be used.
 From the point of view of defining a hamiltonian the previous prescription amounts
to use the density 
\[
\widetilde{\rho }(\vec{r})=\frac{\left\langle \Phi |\hat{\rho }(\vec{r})
|\Phi '\right\rangle }{\left\langle \Phi |\Phi '\right\rangle }
\]
for the density dependent part of the hamiltonian. This prescription coincides with the one obtained
for a Skyrme force with  a linear dependence in $\rho$.  The above density is, in
general, not a real quantity rendering the density dependent ``hamiltonian''
a non hermitian operator. This is not a real problem as the hamiltonian overlap
can be a complex quantity, we only have to be sure that the projected energy
is a real quantity as it will be shown below. 

Another important check concerning
the consistency of the prescription is the following~: since $\hat{P}^{N}$ and the density dependent
hamiltonian do commute  and $\hat{P}^{N^2}=\hat{P}^{N}$, we may write
\begin{equation}
\label{rel1}
E_{N}=\frac{\left\langle \psi _{N}\right| \hat{H}\left| \psi _{N}\right\rangle }{\left\langle
\psi _{N}|\psi _{N}\right\rangle }=\frac{\left\langle \Phi \right| \hat{H}\left| \psi_N
\right\rangle }{\left\langle \Phi \right. \left| \psi_N \right\rangle }.
\end{equation}
According to our prescription, to evaluate  non-diagonal matrix elements of the first
expression of $E_N$ in eq.~(\ref{rel1}), in the hamiltonian we have to consider the 
density 
\begin{equation}
\label{first}
\widetilde{\rho }_{\phi ,\phi '}(\vec{r})=\frac{\left\langle \Phi \right|
e^{-i\phi \hat{N}}\hat{\rho }(\vec{r})e^{i\phi '\hat{N}}\left| \Phi
\right\rangle }{\left\langle \Phi \right| e^{i(\phi '-\phi )\hat{N}}\left| \Phi
\right\rangle },
\end{equation}
whereas for the non-diagonal matrix elements of the second expression of $E_N$
in eq.~(\ref{rel1}) one should use
\begin{equation}
\label{mypres}
\widetilde{\rho }_{\phi }(\vec{r})=
\frac{\left\langle \Phi \right| \hat{\rho }(\vec{r})e^{i\phi \hat{N}}\left| \Phi
\right\rangle }{\left\langle \Phi \right| e^{i\phi \hat{N}}\left| \Phi
\right\rangle }.
\end{equation}
The final result, should be independent of whether we use the first or the second expression. This is clearly
true, since eq.~(\ref{first}) can be written as
\[
\widetilde{\rho }_{0,\phi '-\phi }(\vec{r})=\frac{\left\langle \Phi \right|
\hat{\rho }(\vec{r})e^{i(\phi '-\phi )\hat{N}}\left| \Phi \right\rangle
}{\left\langle \Phi \right| e^{i(\phi '-\phi )\hat{N}}\left| \Phi \right\rangle
},\]
due to the commutativity of \( e^{i\phi \hat{N}} \) with \( \hat{\rho }(\vec{r}) \).
  The operator \( e^{i\phi \hat{N}} \) also commutes with \( \widetilde{\rho }_{0,\phi '-\phi }(\vec{r}) \)
showing that
\begin{eqnarray}
h(\phi ,\phi ') &= &\left\langle \Phi \right| e^{-i\phi \hat{N}}\hat{H}[\widetilde{\rho }_{\phi ,\phi '}(\vec{r})]e^{i\phi '\hat{N}}\left| \Phi \right\rangle 
                 =\left\langle \Phi \right| \hat{H}[\widetilde{\rho }_{0,\phi
		 '-\phi }(\vec{r})]e^{i(\phi '-\phi )\hat{N}}\left| \Phi
		 \right\rangle \nonumber \\
		 &= &h(0,\phi '-\phi ) \nonumber
\end{eqnarray}
 which is the property needed to obtain the second expression of Eq (\ref{rel1}).
Obviously, we shall use the second expression  for the projected energy and 
eq.~(\ref{mypres}),
for the density entering in the density dependent part of the interaction,
in the evaluation of \( h(\phi ) \).

 Next, in order to ensure the reality of the projected energy, we have to show that the following 
 property holds
\begin{equation}
\label{prop1}
h^{*}(\phi )=h(-\phi ).
\end{equation}
 The complex conjugate
of \( \widetilde{\rho }_{\phi }(\vec{r}) \) is given by
\[
\widetilde{\rho }^{*}_{\phi }(\vec{r})=\frac{\left\langle \Phi \right| e^{-i\phi
\hat{N}}\hat{\rho }(\vec{r})\left| \Phi \right\rangle }{\left\langle \Phi
\right| e^{-i\phi \hat{N}}\left| \Phi \right\rangle }=\frac{\left\langle \Phi
\right| \hat{\rho }(\vec{r})e^{-i\phi \hat{N}}\left| \Phi \right\rangle
}{\left\langle \Phi \right| e^{-i\phi \hat{N}}\left| \Phi \right\rangle
}=\widetilde{\rho }_{-\phi }(\vec{r}),\]
 using this property in the definition of \( h^{*}(\phi ) \),
Eq (\ref{prop1}) is demonstrated,
\[
h^{*}(\phi )=\left\langle \Phi \right| e^{-i\phi \hat{N}}\hat{H}(\widetilde{\rho
}^{*}_{\phi }(\vec{r}))\left| \Phi \right\rangle =\left\langle \Phi \right|
\hat{H}(\widetilde{\rho }_{-\phi }(\vec{r}))e^{-i\phi \hat{N}}\left| \Phi
\right\rangle =h(-\phi ).\]

\section{Appendix C: Evaluation of \protect\( \left\langle \Delta \hat{H}(\Delta \hat{N})^{2}\right\rangle \protect \)}

In order to evaluate \( \left\langle \Delta \hat{H}(\Delta \hat{N})^{2}\right\rangle  \)
we will use the following relation
\begin{equation}
\left\langle \Delta \hat{H}(\Delta \hat{N})^{2}\right\rangle =
-\frac{\partial ^{2}}{\partial \phi ^{2}}\left. \frac{\left\langle \Phi \right| \hat{H}e^{i\phi \hat{N}}\left| \Phi \right\rangle }
{\left\langle \Phi \right| e^{i\phi \hat{N}}\left| \Phi \right\rangle
}\right|_{\phi=0}.
\end{equation}
In this expression the density appearing in the density dependent part of the hamiltonian is, obviously
$\langle \Phi | c^+(\vec{r})c(\vec{r}) | \Phi \rangle $. The matrix element
of the right hand side is given by eq.~(\ref{ove}) with $ |\Phi ' \rangle = e^{i\phi \hat{N}}|\Phi \rangle $.
We may further write
\[
\frac{\left\langle \Phi \right| \hat{H}e^{i\phi \hat{N}}\left| \Phi \right\rangle }{\left\langle \Phi \right| e^{i\phi \hat{N}}\left| \Phi \right\rangle }=
Tr\left[(T+\frac{1}{2}\Gamma (\phi ))\widetilde{\rho }(\phi )\right]
-\frac{1}{2}Tr\left[\Delta (\phi )\widetilde{\overline{\kappa }}(\phi )\right],\]
where we have introduced the notation
\[
\Delta_{k_1k_2}(\phi)  =  \frac{1}{2}\sum
_{k_{3}k_{4}}\overline{\nu}_{k_{1}k_{2}k_{3}k_{4}}\widetilde{\kappa}_{k_{3}k_{4}}(\phi),
\;\;\;\;\;
\Gamma_{k_1k_2}(\phi) = \sum
_{k_{3}k_{4}}\overline{\nu}_{k_{1}k_{3}k_{2}k_{4}}\widetilde{\rho
}_{k_{4}k_{3}}(\phi).
\label{dega}
\]
Performing the second derivative with respect to \( \phi  \) and taking the
limit \( \phi =0 \) we obtain
\begin{eqnarray}
\frac{\partial ^{2}}{\partial \phi ^{2}} \left.
\frac{\left\langle \Phi \right| \hat{H}e^{i\phi \hat{N}}\left| \Phi \right\rangle }
{\left\langle \Phi \right| e^{i\phi \hat{N}}\left| \Phi \right\rangle }\right|_{\phi=0}=
&-& Tr\left[T\rho _{2N}+\frac{1}{2}(\Gamma _{2N}\rho +\Gamma \rho _{2N}+2\Gamma _{N}\rho _{N})\right] \nonumber \\
&+& \frac{1}{2}Tr[\Delta _{2N}\kappa ^{*}+\Delta \overline{\kappa }_{2N}+2\Delta
_{N}\overline{\kappa }_{N}],
\end{eqnarray}
 where we have introduced the matrices
\begin{eqnarray*}
\left( \rho _{N}\right) _{k_{1}k_{2}} & = & 
\frac{1}{i}\left. \frac{\partial }{\partial \phi }\widetilde{\rho }_{k_{1}k_{2}}(\phi )\right|_{\phi=0}=
\left\langle \Phi \right| c^{+}_{k_{2}}c_{k_{1}}\Delta \hat{N}\left| \Phi \right\rangle ,\\
\left( \kappa _{N}\right) _{k_{1}k_{2}} & = & 
\frac{1}{i}\left. \frac{\partial }{\partial \phi }\widetilde{\kappa }_{k_{1}k_{2}}(\phi )\right|_{\phi=0}=
\left\langle \Phi \right| c_{k_{2}}c_{k_{1}}\Delta \hat{N}\left| \Phi \right\rangle, \\
\left( \overline{\kappa }_{N}\right) _{k_{1}k_{2}} & = & 
\frac{1}{i}\left. \frac{\partial }{\partial \phi }\widetilde{\overline{\kappa }}_{k_{1}k_{2}}(\phi )\right|_{\phi=0}=
\left\langle \Phi \right| c^{+}_{k_{1}}c^{+}_{k_{2}}\Delta \hat{N}\left| \Phi \right\rangle ,
\end{eqnarray*}
and
\begin{eqnarray*}
\left( \rho _{2N}\right) _{k_{1}k_{2}} & = & 
-\left. \frac{\partial ^{2}}{\partial \phi ^{2}}\widetilde{\rho }_{k_{1}k_{2}}(\phi )\right|_{\phi=0}=
\left\langle \Phi \right| \Delta \left( c^{+}_{k_{2}}c_{k_{1}}\right) \left( \Delta \hat{N}\right) ^{2}\left| \Phi
\right\rangle ,\\
\left( \kappa _{2N}\right) _{k_{1}k_{2}} & = & 
-\left. \frac{\partial ^{2}}{\partial \phi ^{2}}\widetilde{\kappa }_{k_{1}k_{2}}(\phi )\right|_{\phi=0}=
\left\langle \Phi \right| \Delta \left( c_{k_{2}}c_{k_{1}}\right) \left( \Delta \hat{N}\right) ^{2}\left| \Phi
\right\rangle ,\\
\left( \overline{\kappa }_{2N}\right) _{k_{1}k_{2}} & = & 
-\left. \frac{\partial ^{2}}{\partial \phi ^{2}}\widetilde{\overline{\kappa }}_{k_{1}k_{2}}(\phi )\right|_{\phi=0}=
\left\langle \Phi \right| \Delta \left( c^{+}_{k_{1}}c^{+}_{k_{2}}\right) \left( \Delta \hat{N}\right) ^{2}\left| \Phi
\right\rangle, 
\end{eqnarray*}
 $\Gamma_N, \Gamma_{2N}, \Delta_N$ and $\Delta_{2N}$  are given by the usual expressions but the traces have
 to be taken with $\rho_N, \rho_{2N}, \kappa_N$ and $\kappa_{2N}$ respectively.

\end{document}